\documentstyle [a4,12pt,epsf,here,rotate] {article}
\begin{document}

\title{\LARGE {\bf  Relativistic Hartree-Bogoliubov
theory in coordinate space: 
finite element solution for
a nuclear system with spherical symmetry}}

\author{W. P\"oschl\footnotemark[1], D. Vretenar\footnotemark[2], 
and P. Ring\\
Physik-Department der Technischen Universit\"at M\"unchen,\\
D-85748 Garching, Germany}
\vspace{15mm}
\date{\today}
\maketitle
\vspace{15mm}
\footnotetext[1]{Work supported by GSI Darmstadt (GSI TM DIT
36-46-4030) \newline
E-mail:WPOESCHL@PHYSIK.TU-MUENCHEN.DE}
\footnotetext[2]{Alexander von Humboldt Fellow,
on leave of absence from University of Zagreb, Croatia}


%
\abstract{
%
A C++ code for the solution of the relativistic Hartree-Bogoliubov
theory in coordinate space is presented. The theory describes a 
nucleus as a relativistic 
system of baryons and mesons. The RHB model is applied in the 
self-consistent mean-field approximation to the description of 
ground state properties of spherical nuclei. Finite range 
interactions are included to describe pairing correlations and the
coupling to particle continuum states. Finite element methods are used 
in the coordinate space discretization of the coupled system
of Dirac-Hartree-Bogoliubov integro-differential eigenvalue equations,
and Klein-Gordon equations for the meson fields. 
The bisection
method is used in the solution of the  
resulting generalized algebraic eigenvalue problem, and the
biconjugate gradient method for the 
systems of linear and nonlinear algebraic equations, respectively.
}
%
%
%

\newpage
\centerline{\Large \bf PROGRAM SUMMARY}
\vspace{1 cm}
\begin{itemize}
\item[]{\it Title of program\/}:~ spnRHBfem.cc\hfill\break
\item[]{\it Catalogue number\/}:~..........\hfill\break
\item[]{\it Program obtainable from\/}:~   \hfill\break
\item[]{\it Computer for which the program is designed and others 
on which it has been tested\/}:~any Unix work-station. \hfill\break
\item[]{\it Operating system\/}:~Unix \hfill\break
\item[]{\it Programming language used\/}:~C++ \hfill\break
\item[]{\it No. of lines in combined program and test 
deck\/}:                                       \hfill\break
\item[]{\it Keywords\/}:~ relativistic Hartree-Bogoliubov theory, 
mean-field approximation, spherical nuclei, pairing, 
Dirac-Hartree-Bogoliubov equations, Klein-Gordon 
equation, Finite Element Method,
bisection method, classes \hfill\break \vskip 0.5cm
\item[]{\it Nature of physical problem}\hfill\break
The ground-state of a spherical nucleus is 
described in the framework of relativistic Hartree-Bogoliubov 
theory in coordinate space. The model describes a
nucleus as a relativistic system of baryons and mesons. 
Nucleons interact in a relativistic covariant manner 
through the exchange of virtual mesons:~the isoscalar scalar $\sigma$-meson, 
the isoscalar vector $\omega$-meson and the isovector vector $\rho$-meson. 
The model is based on the one boson exchange description of the 
nucleon-nucleon 
interaction. Pairing correlations are described by finite range 
Gogny forces.
\\
\item[]{\it Method of solution}\hfill\break
An atomic nucleus is described by a coupled system of partial 
integro-differential
equations for the nucleons (Dirac-Hartree-Bogoliubov equations),
and differential equations for the meson and photon
fields (Klein-Gordon equations). A method is presented which allows
a simple, self-consistent solution based on finite element analysis.
Using a formulation based on weighted residuals, the coupled system
of Dirac-Hartree-Bogoliubov 
and Klein-Gordon equations is transformed into a generalized
algebraic eigenvalue problem, and systems of linear and nonlinear  
algebraic equations, respectively. Finite
elements of arbitrary order
are used on adaptive non-uniform radial mesh.
The generalized eigenvalue problem is solved in narrow windows of the
eigenparameter using a highly efficient bisection method for band matrices.
A biconjugate gradient method is used for the solution of systems of linear
and nonlinear algebraic equations. 

\item[]{\it Restrictions on the complexity of the problem}\newline
In the present version of the code we only consider nuclear systems 
with spherical symmetry. 

\end{itemize}
\bigskip

\centerline{\large \bf LONG WRITE-UP}
%
\section {Introduction}
\noindent 
Relativistic mean-field theory has been extensively applied in 
calculations of nuclear matter and properties of finite nuclei throughout
the periodic table. The theory provides a framework for 
describing the nuclear many-body problem as a relativistic 
system of baryons and mesons~\cite{SW.86,Rei.89,Ser.92}. 
In the self-consistent mean-field 
approximation, detailed calculations have been performed for a variety
of nuclear stucture phenomena (for a recent review see ~\cite{Rin.96}).
More recently, the 
relativistic mean-field model has been also applied in the 
description of the structure of exotic nuclei with extreme isospin 
values. This new field includes many interesting phenomena:
extremely weak binding of the outermost 
nucleons, coupling between bound states and the particle continuum, 
regions of neutron halos with very diffuse neutron densities,  
large spatial dimensions and the existence of the neutron skin.
Major modifications of shell structures in drip-line nuclei
have been predicted, as well as modifications in the
onset and evolution of collectivity. For most phenomena 
along the line of stability, non-relativistic models and the
relativistic framework predict very similar results, although
relativistic mean-field models provide a more economical 
description. For drip-line nuclei, on the other hand, one 
also expects differences in the predictions of non-relativistic 
and relativistic models, especially in the treatment of the 
spin-orbit interaction \cite{Rin.96}.

In drip-line nuclei the Fermi level is found close to the 
particle continuum. The lowest particle-hole or particle-particle 
modes are often embedded in the continuum, and the coupling between 
bound and continuum states has to be taken into account explicitly. 
In the mean-field approximation the most important residual interaction 
is the pairing force. Therefore, in the description of 
ground-state properties, it is essential to include 
mean-field and pairing correlations simultaneously. 
The Relativistic Hartree-Bogoliubov (RHB) theory in coordinate
space, which is an extension of non-relativistic
HFB-theory~\cite{DFT.84}, provides such a unified description. 
In particular, it includes the scattering of nucleonic pairs from 
bound states to the positive energy continuum.
The RHB theory has recently been applied in the description
of ground-state properties of Sn and Pb isotopes~\cite{LEL.96}, using 
an expansion in a large oscillator basis for the 
solution of the Dirac-Hartree-Bogoliubov equations.
In many applications an expansion of the wave functions in an 
appropriate oscillator basis of spherical or axial symmetry provides
a satisfactory level of accuracy.  
However, in the case of drip-line nuclei, the expansion in the localized
oscillator basis presents only a poor approximation to 
the continuum states, and the  convergence of such expansions is 
too slow and is not uniform. Examples are exotic phenomena such 
as neutron halos and neutron skins. In order to correctly
describe the coupling between bound and 
continuum states, the Dirac-Hartree-Bogoliubov equations have to be 
solved in coordinate space. Discretization in coordinate space provides
also the advantage that exotic shapes and/or large deformations 
can be described, without preparing a basis specific to each 
deformation. Recently, a fully self-consistent RHB model in 
coordinate space has been used to describe the two-neutron halo 
in $^{11}$Li~\cite{MR.96}. However, only a density dependent 
force of zero range has been used in the pairing channel. 
In general, it is assumed that a finite range interaction would 
provide a more realistic description of pairing correlations. 
In the present article we describe a C++ code which can be 
used to calculate ground state properties of spherical nuclei, 
within the framework of RHB theory in coordinate space, with 
finite range forces in the pairing channel. 

A convenient procedure for the coordinate space discretization 
of the Dirac-Hartree-Bogoliubov equations is 
provided by Finite Element 
Methods (FEM)~\cite{Zink.77,Hint.79,Jin.93,Gros.94}.
In Refs.~\cite{PVR.96,PVR.97} we have applied FEM to the solution of the 
coupled system of relativistic mean-field equations in the description of 
a one-dimensional slab of nuclear matter and of spherical 
doubly closed-shell nuclei. We have investigated 
the applicability of FEM in the calculation of 
bound and continuum eigenstates of the Dirac equation.   
Since the spectrum of the hamiltonian of the 
hyperbolic Dirac equation is not bounded from below, finite 
element methods cannot be applied in the variational formulation. 
The method of weighted residuals produces element matrix integral 
definitions that would be identical to those obtained from a 
variational form, if one existed. Our analysis has shown that
FEM provide very accurate solutions for the relativistic 
eigenvalue problem in the self-consistent mean-field approximation.
In the present work we extend the model to include 
pairing correlations in the framework of RHB theory. 

The article is organized as follows. In Sec. 2 the Relativistic 
Hartree-Bogoliubov theory is described. Dirac-Hartree-Bogoliubov 
equations for a system with spherical symmetry are derived. 
The finite element analysis is described in Sec. 3. In Sec. 4 
we present some illustrative calculations and discuss the quality of our 
approximations and numerical results. The structure of the 
C++ code is described in Sec. 5. 
%
%
%
%
\section {The relativistic Hartree-Bogoliubov equations}
%
The Hartree-Fock-Bogoliubov (HFB) theory provides a unified description of 
mean-field and pairing correlations in nuclei~\cite{RISH.80}. 
Independent quasiparticles are introduced and the ground state 
of a nucleus $\vert \Phi >$ is represented as the vacuum with 
respect to these quasi-particles. The quasi-particle operators 
are defined by a unitary Bogoliubov transformation of the 
single-nucleon creation and annihilation operators. 
The generalized single-particle hamiltonian of HFB theory
contains two average potentials: the self-consistent field 
$\hat\Gamma$ which encloses all the long range {\it ph} correlations,
and a pairing field $\hat\Delta$ which sums up the 
{\it pp}-correlations. The expectation value of the nuclear hamiltonian
$< \Phi\vert \hat H \vert \Phi >$ can be expressed as a function of the 
hermitian density matrix $\rho$, and the antisymmetric pairing 
tensor $\kappa$. The variation of the energy functional with respect
to $\rho$ and $\kappa$ produces the single quasi-particle 
Hartree-Fock-Bogoliubov equations (for details of the derivation 
we refer to~\cite{RISH.80}).
\begin{eqnarray}
\label{equ.2.1}
\left( \matrix{ \hat h - \lambda & \hat\Delta \cr
                -\hat\Delta^* & -\hat h +\lambda
                 } \right) \left( \matrix{ U_k \cr V_k}\right) =
E_k\left( \matrix{ U_k \cr V_k } \right).
\end{eqnarray}
HFB-theory, being a variational approximation, results in a 
violation of basic symmetries of the nuclear system, among which 
the most important is the nonconservation of the number of particles.
In order that the expectation value of the particle number operator in the
ground state equals the number of nucleons,
equations (\ref{equ.2.1}) contain a chemical potential 
$\lambda$ which has to be determined by the particle number subsidiary 
condition. The column vectors denote the quasi-particle 
wave functions, and $E_k$ are the quasi-particle energies. 

The relativistic extension of the HFB theory is descibed in Ref.~ 
\cite{KR.91}. In the Hartree approximation for the self-consistent 
mean field, the Relativistic Hartree-Bogoliubov (RHB) equations read
\begin{eqnarray}
\label{equ.2.2}
\left( \matrix{ \hat h_D -m- \lambda & \hat\Delta \cr
                -\hat\Delta^* & -\hat h_D + m +\lambda
                 } \right) \left( \matrix{ U_k \cr V_k } \right) =
E_k\left( \matrix{ U_k \cr V_k } \right).
\end{eqnarray}
where $\hat h_D$ is the single-nucleon Dirac hamiltonian~\cite{PVR.97},
and $m$ is the nucleon mass. The RHB equations are non-linear 
integro-differential equations. They have to be solved self-consistently, 
with potentials determined in the mean-field approximation from 
solutions of Klein-Gordon equations for mesons~\cite{PVR.97}
\begin{eqnarray}
\label{equ.2.3.a}
\bigl[-\Delta + m_{\sigma}^2\bigr]\,\sigma({\bf r})&=&
-g_{\sigma}\,\rho_s({\bf r})
-g_2\,\sigma^2({\bf r})-g_3\,\sigma^3({\bf r})   \\
\label{equ.2.3.b}
\bigl[-\Delta + m_{\omega}^2\bigr]\,\omega^0({\bf r})&=&
-g_{\omega}\,\rho_v({\bf r}) \\
\label{equ.2.3.c}
\bigl[-\Delta + m_{\rho}^2\bigr]\,\rho^0({\bf r})&=&
-g_{\rho}\,\rho_3({\bf r}) \\
\label{equ.2.3.d}
-\Delta \, A^0({\bf r})&=&e\,\rho_p({\bf r}).
\end{eqnarray}
for the sigma meson, omega meson, rho meson and photon field, respectively.
The spatial componenets $\bf\omega$, $\bf\rho$, and {\bf A} vanish 
due to time reversal symmetry. Because of charge conservation, only the 
3-component of the isovector rho meson contributes. The source terms in 
equations (\ref{equ.2.3.a}) to (\ref{equ.2.3.d}) are sums of bilinear 
products of baryon amplitudes 
\begin{eqnarray}
\label{equ.2.3.e}
\rho_s&=&\sum\limits_{E_k > 0} V_k^{\dagger}\gamma^0 V_k, \\
\label{equ.2.3.f}
\rho_v&=&\sum\limits_{E_k > 0} V_k^{\dagger} V_k, \\
\label{equ.2.3.g}
\rho_3&=&\sum\limits_{E_k > 0} V_k^{\dagger}\tau_3 V_k, \\
\label{equ.2.3.h}
\rho_{\rm em}&=&\sum\limits_{E_k > 0} V_k^{\dagger} {{1-\tau_3}\over 2} V_k.
\end{eqnarray}
where the sums run over all positive energy states. For M degrees of 
freedom, for example number of nodes on a radial mesh, the HB equations
are 2M-dimensional and have 2M eigenvalues and eigenvectors. To each
eigenvector $(U_k, V_k)$ with eigenvalue $E_k$, there corresponds an
eigenvector $(V^*_k, U^*_k)$ with eigenvalue $-E_k$. Since baryon 
quasi-particle operators satisfy fermion commutation relations, it is 
forbidden to occupy the levels $E_k$ and $-E_k$ simultaneously. Usually 
one chooses the M positive eigenvalues $E_k$ for the solution that 
corresponds to a ground state of a nucleus with even particle number.

The system of equations
(\ref{equ.2.2}), and (\ref{equ.2.3.a}) to (\ref{equ.2.3.d}), 
is solved self-consistently in coordinate space
by discretization on the finite element mesh. 
In the coordinate space representation of the pairing field 
$\hat\Delta $ in (\ref{equ.2.2}), the kernel of the integral operator is
\begin{equation}
\label{equ.2.5}
\Delta_{ab} ({\bf r}, {\bf r}') = {1\over 2}\sum\limits_{c,d}
V_{abcd}({\bf r},{\bf r}') {\bf\kappa}_{cd}({\bf r},{\bf r}').
\end{equation}
where $a,b,c,d$ denote all quantum numbers, apart from the coordinate
$\bf r$, that specify the single-nucleon states.
$V_{abcd}({\bf r},{\bf r}')$ are matrix elements of a general 
two-body pairing interaction, and the pairing tensor is defined as
\begin{equation}
{\bf\kappa}_{cd}({\bf r},{\bf r}') := 
\sum_{E_k>0} U_{ck}^*({\bf r})V_{dk}({\bf r}').
\end{equation}
The integral operator $\hat\Delta$ acts on the wave function
$V_k({\bf r})$:
\begin{equation}
\label{equ.2.4}
(\hat\Delta V_k)({\bf r}) 
= \sum_b \int d^3r' \Delta_{ab} ({\bf r},{\bf r}') V_{bk}({\bf r}'). 
\end{equation}

The eigensolutions of Eq. (\ref{equ.2.2}) form a set of
orthogonal (normalized) single quasi-particle states. The corresponding
eigenvalues are the single quasi-particle energies.
The Bogoliubov transformation
from the single-particle coordinate basis of $\delta$-functions 
to the basis of quasi-particle
states is given by the matrix 
$W_{k{\bf r}}:= (U_k^T({\bf r}),V_k^T({\bf r}))^T $.
${\bf r}$ and $k$ are column and row indices, respectively. 
In the self-consistent iteration procedure we work 
in the basis of quasi-particle states. The self-consistent quasi-particle
eigenspectrum is then transformed into the canonical basis of single-particle
states. The canonical basis is defined to be the one 
in which the matrix 
$R_{kk'}=\bigl< V_k({\bf r})\big\vert V_{k'}({\bf r})\bigl>$ is diagonal.
The transformation to the canonical basis determines the energies
and occupation probabilities of single-particle states, that correspond 
to the self-consistent solution for the ground state of a nucleus.
In order to determine the canonical basis, we have two possibilities:
either diagonalize the density matrix,
or diagonalize the matrix
\begin{equation}
R_{kk'}~:=~\bigl< V_k({\bf r})\big\vert V_{k'}({\bf r})\bigl> 
\end{equation}
Although both methods
are in principle equivalent, for numerical reasons we chose
the second method, i.e. we diagonalize the matrix $R_{kk'}$.
Because of the truncation in quasiparticle space, 
the dimension of the matrix $R_{kk'}$,
which is a matrix in quasiparticle space, is considerably
smaller than the dimension of the density matrix 
in coordinate space.

The transformations from the general single-particle basis to the basis
of quasi-particle states, and the canonical basis are illustrated 
in Fig. 1. The quasi-particle operators and basis states are defined
\begin{equation}
\label{equ.2.5.a}
\pmatrix{ \hat a_k \cr \hat a_k^{\dagger} } = 
\int d^3 r \pmatrix{ U^*_{k,{\bf r}} & V^*_{k,{\bf r}} \cr
                     V_{k,{\bf r}}   & U_{k,{\bf r}} }
\pmatrix{ \hat\Psi({\bf r}) \cr \hat\Psi^{\dagger }({\bf r}) }, \qquad
\pmatrix{ \Phi_U({\bf r}) \cr \Phi_V({\bf r}) }\,\,
=\,\, \int\limits_{{\bf R}^3}
d^3r' \left( \matrix{ U_{k,{\bf r}'} \cr V_{k,{\bf r}'} } \right)
\delta ({\bf r}-{\bf r}'). 
\end{equation}
The operators in the canonical basis are 
\begin{equation}
\label{equ.2.5.b}
\pmatrix{ \hat c_k \cr \hat c_k^{\dagger } } = 
\int d^3 r \pmatrix{ \psi_k({\bf r}) & 0                     \cr
                     0               & \psi^*_k({\bf r}) }
\pmatrix{ \hat\Psi({\bf r})\cr\hat\Psi^{\dagger}({\bf r})},\quad\qquad 
\psi_k({\bf r})\,\, =\,\, \int\limits_{{\bf R}^3}d^3r'\Psi_{k,{\bf r}' }
\delta ({\bf r}-{\bf r}'), 
\end{equation}
and the transformation from the quasi-particle to the canonical basis reads
\begin{equation}
\label{equ.2.5.c}
\pmatrix{ \hat c_k \cr \hat c_k^{\dagger} } = 
\sum\limits_{k'} \pmatrix{ u_{kk'} & v_{kk'}                     \cr
                          -v_{kk'} & u_{kk'} }
\pmatrix{ \hat a_{k'}\cr\hat a^{\dagger}_{k'} }, \qquad\qquad\quad 
\psi_k({\bf r})\,\, =\,\, \sum\limits_{k'} C_{k'k}\Phi_{V,k'}({\bf r})
\end{equation}
The matrix representation of the unitary 
Bogoliubov transformation ${\cal{\bf W}}$ can be 
decomposed into a product of three matrices~\cite{RISH.80} 
%
%
%
%
\begin{eqnarray}
\label{equ.2.6}
{\cal{\bf W}} = \pmatrix{ {\bf D} & 0 \cr 0 & {\bf D}^* }
         \pmatrix{ \bar {\bf U} & \bar {\bf V} \cr 
                  - \bar {\bf V} & \bar {\bf U} }
         \pmatrix{ {\bf C} & 0 \cr 0 & {\bf C}^* }.
\end{eqnarray}
The diagonalization of the matrix 
$R_{kk'}:=\bigl< V_k({\bf r})\vert V_{k'}({\bf r})\bigr>$ 
produces the unitary matrix ${\bf C}$.
Columns of ${\bf C}$ are eigenvectors of $R_{kk'}$. 
The matrices $\bar {\bf U}$ and $\bar {\bf V}$ are constructed 
from the eigenvalues
$v_k^2$ of $R_{kk'}$
\begin{eqnarray}
\label{equ.2.7}                                 
\bar {\bf U}=\pmatrix{ u_1 & 0 & . & . & . &   & 0 \cr 
                 0   & . &   &   &   &   &   \cr
                 .   &   & . &   &   &   & . \cr
                 .   &   &   &u_k&   &   & . \cr
                 .   &   &   &   & . &   & . \cr
                     &   &   &   &   & . & 0 \cr 
                 0   &   & . & . & .. & 0 & u_N }
\qquad \mbox{\rm and} \qquad
\bar {\bf V}=\pmatrix{ v_1 & 0 & . & . & . &   & 0 \cr 
                 0   & . &   &   &   &   &   \cr
                 .   &   & . &   &   &   & . \cr
                 .   &   &   &v_k&   &   & . \cr
                 .   &   &   &   & . &   & . \cr
                     &   &   &   &   & . & 0 \cr 
                 0   &   & . & . & . & 0 & v_N }
\end{eqnarray}
where $u_k = \sqrt{1-v_k^2}$, and $N$ is the number of 
solutions.
The matrix ${\bf C}$ transforms
the basis of quasi-particle states into the canonical basis 
\begin{equation}
\psi_k({\bf r}) =\sum\limits_{k'}C_{k'k}V_{k'}({\bf r}).
\end{equation}
The matrix ${\bf D}:=\left[\psi_k({\bf r})\right]$ 
represents a transformation between two single-particle bases. 
${\bf E}:={\rm diag}(E_k)$ defines the diagonal matrix of quasi-particle
energies.
In the basis of quasi-particle states the
hamiltonian matrix has the diagonal form 
${\rm diag}( {\bf E},-{\bf E})$. 
The hamiltonian matrix in the single-particle canonical basis 
is defined by the transformation
\begin{eqnarray}
\label{equ.2.8}
{\bf H}:=\pmatrix{ \bar{\bf U} & \bar{\bf V} \cr
                  -\bar{\bf V} & \bar{\bf U} }
        \pmatrix{ {\bf C} & 0 \cr
                        0   & {\bf C}^* }
        \pmatrix{ {\bf E} & 0 \cr
                    0     & -{\bf E} }
        \pmatrix{ {\bf C}^{\dagger} & 0 \cr
                   0      & {\bf C}^T }
\pmatrix{ \bar{\bf U} & -\bar{\bf V} \cr
                 \bar{\bf V}^{\dagger} & \bar{\bf U}^{\dagger} }.
\end{eqnarray}
The single-particle energies correspond to the diagonal matrix elements
$\varepsilon_n = H_{nn} + \lambda$, where $\lambda$ denotes the chemical
potential.
%
%
%
%
\noindent
In Fig. 2 we display a schematic single-nucleon spectrum in the 
relativistic mean-field potential of a finite nucleus. On the left
hand side the eigenspectrum of a Dirac hamiltonian is shown. 
The single-particle hamiltonian corresponds to the average 
mean-field potential, and the Dirac equation is solved in the
Hartree mean-field, and {\it no-sea} approximations. 
A Dirac gap is observed between states of negative and
positive energy. 
In vacuum, this gap equals two times the nucleon mass. 
In a nucleus the Dirac gap extends 
between the sum and the difference of 
the scalar sigma-meson potential and vector omega-meson potential. The 
sum and the difference are given relative to $+m$ and $-m$, 
respectively. In the center of Fig. 2 the eigenspectrum 
of the Dirac hamiltonian is shifted by the nucleon mass $m$. As a
result, bound single-nucleon levels have negative energies, while
the positive energy domain contains only single-nucleon continuum states.
The single quasi-particle spectrum which results as a solution of
the relativistic Hartree-Bogoliubov equations is shown on the 
right hand side of Fig. 2. The number of solutions is two times
the number of physical states. As already described, for each
eigenvector $(U_k, V_k)$ with energy $E_k$, the
corresponding state $(V^*_k, U^*_k)$ is found at $-E_k$. 

In practical calculations the Dirac-Hartree-Bogoliubov and
Klein-Gordon equations are discretized on a finite domain
${\cal D}$ in coordinate space (indicated by $r_{max}$ in Fig. 2), 
and the generalized eigenvalue problem is solved in the window 
${\cal E} := [0,E_{max}]$ of the eigenparameter E. 
The domain
${\cal D}\otimes {\cal E}$ is indicated by the shaded area in the 
right hand side spectrum. By increasing the coordinate space domain 
the HB spectrum becomes denser, and thus provides a better 
approximation for the continuum. Larger values of $E_{max}$ take 
into account couplings to highly excited quasi-particle states. 
${\cal D}$ and ${\cal E}$ should be chosen in such a way that the 
resulting densities do not depend on their precise values. In particular,
$E_{max}$ has to be larger than the absolute value of the depth 
of the potential well. 

In the present version of the code we only consider single closed-shell
nuclei, i.e. systems with spherical symmetry. 
The fields $\sigma(r),\,\omega^0(r),\,
\rho^0(r),$ and $A^0(r)$ depend only on the radial coordinate $r$. 
The nucleon spinors $U_k$ ($V_k$) in (\ref{equ.2.2}) are characterized 
by the angular
momentum $j$, its $z$-projection $m$, parity $\pi$ and the isospin
$t_3=\pm {1\over 2}$ for neutron and proton. We combine the two
Dirac spinors $U_k({\bf r})$ and $V_k({\bf r})$ to form a {\it super-spinor}
\begin{equation}
\Phi_k({\bf r})~:=~\pmatrix{ U_k({\bf r}) \cr V_k({\bf r})\cr}
\end{equation}
where
\begin{eqnarray}
\label{spherspinor}
{U_k(V_k)}({\bf r},s,t_3)=
\pmatrix{ g_{U(V)}(r)\Omega_{j,l,m} (\theta,\varphi,s) \cr
        if_{U(V)}(r)\Omega_{j,\tilde l,m} (\theta,\varphi,s) \cr }
         \chi_\tau(t_{3}).
\end{eqnarray}
$g(r)$ and $f(r)$ are radial amplitudes,
$\chi_\tau$ is the isospin function, the
orbital angular momenta $l$ and $\tilde l$ are determined by $j$ and
the parity $\pi$
\begin{eqnarray}
l=\left\{ \matrix{ j+1/2 & \mbox{ for} & \pi = (-1)^{j+1/2} \cr
                  j-1/2 & \mbox{ for} & \pi = (-1)^{j-1/2} \cr } \right., 
\end{eqnarray}
and
\begin{eqnarray}
\tilde l=\left\{ \matrix{ j-1/2 & \mbox{ for} & \pi = (-1)^{j+1/2} \cr
                   j+1/2 & \mbox{ for} & \pi = (-1)^{j-1/2} \cr } \right. 
\end{eqnarray}
$\Omega_{jlm}$ is the tensor product of the orbital and spin functions
\begin{equation}
\Omega_{j,l,m} (\theta,\varphi,s)=\sum\limits_{m_s,m_l}
\bigl< {1\over 2}m_slm_l\big\vert jm\bigr> 
\chi_{{1\over 2} m_s} Y_{lm_l}(\theta,\varphi).
\end{equation}
It will be useful to define a single angular quantum number $\kappa$
as the eigenvalue of the operator $(1+\hat{\bf\sigma}\cdot{\hat{\bf l}}\,)$
\begin{equation}
\label{equ.2.11}
(1+\hat{\bf\sigma}\cdot{\hat{\bf l}}\,)\,\Omega_{\kappa,m} 
= -\kappa\,\Omega_{\kappa,m},
\end{equation}
\begin{equation}
\kappa=\pm(j+{1/2})~~~~ {\rm for}~~~~ j=l\mp{1/2}.
\end{equation}
$\kappa = \pm 1,\pm 2,\pm 3,...$, and the Dirac HB equations are solved 
in coordinate space for each value of $\kappa$.
The equations for the radial amplitudes $g_{U(V)}(r)$ and $f_{U(V)}(r)$ 
are derived from Eq. (\ref{equ.2.2}). The radial single-particle 
Dirac Hamiltonian reads (see also appendix A)
\begin{equation}
\label{equ.2.13}
\hat h_{\kappa}(r)-m:=
\sigma_3\sigma_1 (\partial_r+r^{-1})-\sigma_1\kappa r^{-1}+
(\sigma_3-{\bf 1}_2)m+\sigma_3 S(r)+{\bf 1}_2 V_0(r),
\end{equation}
where the scalar and vector potentials are
\begin{equation}
\label{scapot}
S(r) = g_\sigma \sigma(r),~~~ {\rm and}~~~
V^0(r)=g_{\omega}\,\omega^0(r)+
g_{\rho}\,\tau_3\,\rho^0_3(r)
+e{{(1-\tau_3)}\over 2}A^0(r).
\end{equation}
$\sigma_i$ (i=1,2,3) are the Pauli matrices. The integral
operator of the pairing interaction takes the form
\begin{equation}
\label{equ.2.14}
\hat\Delta (r) =\int\limits_0^{\infty}dr'{r'}^2 \Delta(r,r') 
\end{equation}
The kernel of the integral operator is defined 
\begin{equation}
\label{equ.2.15}
\Delta_{aa'}^{JM}(r,r')={1\over 2}\sum_{\tilde a,\tilde a'}
\bigl< r a,r' a'\big\vert
\hat V\big\vert r\tilde a,r'\tilde a'\bigr>_{JM}
{\bf\kappa}_{\tilde a,\tilde a'}(r,r')
\end{equation}
where $r$ and $r'$ denote radial coordinates, $a$, $a'$, $\tilde a$ and 
$\tilde a'$ are quantum numbers that completely specify single-nucleon 
states: $(n,l,j,m)$ or $(n,\kappa,m)$,
$J$ and $M$ are the total angular momentum of the pair, and its $z$-projection,
respectively. For the pairing interaction we use the finite range Gogny force
\begin{equation}
\label{equ.2.18}
V^{pp}(1,2)=\sum_{i=1,2} e^{-{ {({\bf r}_1 -{\bf r}_2)^2 }\over{\mu_i^2}} }
(W_i +B_i P^\sigma -H_i P^\tau-M_i P^\sigma P^\tau).
\end{equation}
We only consider contributions from $J=0$ pairs to the pairing matrix
elements.
The kernel of the integral operator can then be written
\begin{equation}
\label{equ.2.20}
\Delta_{\kappa} (r,r') = {1\over 2}\sum\limits_{\tilde\kappa}
V^{J=0}_{\kappa\tilde\kappa}(r,r') {\bf\kappa}_{\tilde\kappa}( r, r').
\end{equation}
Details on the calculation of two-body matrix elements of the Gogny 
force are given in Appendix B.
The pairing tensor is calculated 
\begin{equation}
\label{equ.2.21}
{\bf\kappa}_{\kappa}(r,r')=\sum\limits_n 2\vert\kappa\vert
\pmatrix{
g^{(U)}_{n,\kappa}(r)g_{n,\kappa}^{(V)}(r') & 0 \cr
  0 & f^{(U)}_{n,\kappa}(r)f_{n,\kappa}^{(V)}(r') }
\end{equation}
where, for a quantum number $\kappa$,  the sum runs over all solutions
in the specified energy interval $0 < E < E_{\rm max}$.
If we define $\Phi_{U(V)}(r):=(g_{U(V)}(r),f_{U(V)}(r))^T$, the radial 
Dirac-Hartree-Bogoliubov equations read
\begin{eqnarray}
\label{equ.2.22}
(\hat h_D(r)-m -\lambda )\Phi_U(r)+\int_0^{\infty}dr'r'^2\Delta(r,r')\Phi_V(r') 
= E\Phi_U(r) \nonumber \\
(-\hat h_D(r)+m +\lambda )\Phi_V(r)+\int_0^{\infty}dr'r'^2\Delta(r,r')\Phi_U(r') 
= E\Phi_V(r)
\end{eqnarray}
The meson and photon fields are solution of the Klein-Gordon equations
\begin{eqnarray}
\label{equ.2.23.a}
\bigl(-\partial_r^2 - {2\over r}\partial_r +{{l(l+1)\over r^2}}
+ m_{\sigma}^2\bigr)\,\sigma(r)&=&-g_{\sigma}\,\rho_s(r)
-g_2\,\sigma^2(r)-g_3\,\sigma^3(r)   \\
\label{equ.2.23.b}
\bigl(-\partial_r^2 - {2\over r}\partial_r +{{l(l+1)\over r^2}}
+ m_{\omega}^2\bigr)\,\omega^0(r)&=&
-g_{\omega}\,\rho_v(r) \\
\label{equ.2.23.c}
\bigl(-\partial_r^2 - {2\over r}\partial_r +{{l(l+1)\over r^2}}
+ m_{\rho}^2\bigr)\,\rho^0(r)&=&-g_{\rho}\,\rho_3(r) \\
\label{equ.2.23.d}
\bigl(-\partial_r^2 - {2\over r}\partial_r +{{l(l+1)\over r^2}} \bigr)
\, A^0(r)&=& e\,\rho_p(r).
\end{eqnarray}
where $l=0$ for spherical nuclei. In the following we describe the 
computer code used to solve the system of equations 
(\ref{equ.2.22}) to (\ref{equ.2.23.d})
in a self-consistent iteration scheme.
\vskip 1.5cm
%
%
%
%
\section {Finite Element discretization of the radial equations}
%

The radial
equations (\ref{equ.2.22}) - (\ref{equ.2.23.d}) can be written in
the general form
\begin{eqnarray}
\label{radsys1}
\hat H(\lambda,r,\sigma(r),\omega^0(r),\rho^0(r),A^0(r),\Delta(r),
\lambda)\Phi_i(r)&=&
\varepsilon_i\Phi_i(r)
\, \\
\label{radsys2}
\hat R_{\sigma}(r,\sigma(r))\sigma(r) &=& s_{\sigma}(\Phi_1(r),...,\Phi_A(r))
\, \\
\label{radsys3}
\hat R_{\omega}(r)\omega^0(r)&=&s_{\omega}(\Phi_1(r),...,\Phi_A(r))
\, \\
\label{radsys4}
\hat R_{\rho}(r)\rho^0(r)&=&s_{\rho}(\Phi_1(r),...,\Phi_A(r))
\, \\
\label{radsys5}
\hat R_{C}(r)A^0(r) &=& s_{C}(\Phi_1(r),...,\Phi_Z(r)).
\end{eqnarray}
where $\hat H$ is a $4\times 4$ matrix operator defined as
\begin{eqnarray}
\label{radHam}
\hat H:= \nonumber \\
\left( \matrix{ 
       S(r)+V(r)-\lambda & \partial_r-(\kappa_i -1) r^{-1} 
       & \hat\Delta^{(gg)}(r) & 0                      \cr 
      -\partial_r-(\kappa_i +1) r^{-1} & -2m-S(r)+V(r)-\lambda & 0
       & \hat\Delta^{(ff)}(r)                           \cr
       \hat\Delta^{(gg)}(r) & 0 &
       -S(r)-V(r)+\lambda & -\partial_r+(\kappa_i -1) r^{-1} \cr 
       0  & \hat\Delta^{(ff)}(r) &                          
      \partial_r+(\kappa_i +1) r^{-1} & 2m+S(r)-V(r)+\lambda \cr
      } \right).
\end{eqnarray}
and
\begin{eqnarray}
\label{radMesOp1}
\hat R_{\sigma} &:=&
-\partial_r^2-2\, r^{-1}\,\partial_r+ l(l+1)\,r^{-2}+
m_{\sigma}^2+g_2\,\sigma(r)+g_3\,\sigma^2(r)  \, \\
\label{radMesOp2}
\hat R_{\omega} &:=&
-\partial_r^2-2\, r^{-1}\,\partial_r+l(l+1)\, r^{-2}+
m_{\omega}^2 \, \\
\label{radMesOp3}
\hat R_{\rho} &:=&
-\partial_r^2-2\, r{-1}\partial_r+l(l+1)\, r^{-2}+
m_{\rho}^2 \, \\
\label{radMesOp4}
\hat R_C &:=&
-\partial_r^2-2\, r^{-1}\partial_r+l(l+1)\,r^{-2}.
\end{eqnarray}
The source terms on the right hand sides of these equations are defined
as $s_{\sigma}:=-g_{\sigma}\,\rho_{\rm s}(r)$, 
$s_{\omega}:=g_{\omega}\,\rho_{\rm v}(r)$, $s_{\rho}:=g_{\rho}\,\rho_3(r)$,
and $s_C:=e\,\rho_{\rm em}(r)$. The method of finite elements is used
to discretize the system of Dirac-Hartree-Bogoliubov and Klein-Gordon 
equations on a spherical mesh. For the nucleon spinor we 
define $\Psi(r):=(\Phi_U(r)^T,\Phi_V(r)^T)^T$, and use
the FEM ansatz
\begin{equation} 
\label{equ.3.1}
\Psi(r)=\sum\limits_p X_p N_p(r)\qquad (X_p\epsilon{\bf R}^4),
\end{equation}
where the coefficients $X_p$ are four-component vectors,
and  $N_p(r)$ denotes Lagrange shape functions of arbitrary
order~\cite{PVR.97}. 
The index $p$ enumerates nodes on a finite element mesh
for a spherical box 
($r_{\rm min}=0$ to $r_{\rm max}$).
$r_p$ denotes the radial coordinate of the node $p$.
The Lagrange shape functions $N_p$ satisfy the property
$N_p(r_{p'})=\delta_{pp'}$. Therefrom the coefficients $X_p$ in the ansatz 
(\ref{equ.3.1}) correspond to the amplitudes $\Psi(r_p)$ 
of the solution: $X_p=(g^{(U)}(r_p),f^{(U)}(r_p),
g^{(V)}(r_p),f^{(V)}(r_p))^T$. 
The Dirac-Hartree-Bogoliubov equations can be formally written
\begin{equation}
\label{Eq.3.6}
\hat D(\Psi) = f(\Psi,x)
\end{equation}
where $\hat D$ is a differential operator. For any approximate solution
$\tilde\Psi(x)$, the residual error is defined as
\begin{equation}
\label{Eq.3.7}
R(x) := \hat D(\tilde\Psi) - f(\tilde\Psi,x).
\end{equation}
If (\ref{Eq.3.6}) has to be solved on a compact domain 
$\Omega\subset {\bf R}^n$,
where $\Psi(x)=g(x)$ for $x\,\epsilon\,\partial\Omega$, and $g(x)$ is defined
on the boundary $\partial\Omega$ of $\Omega$, the method of 
weighted residuals can be used to define the 
weak form of the boundary value problem 
\begin{equation}
\int\limits_{\Omega}R(x)\, w(x)\, d^nx \equiv 0\quad 
{\rm for\, all\, weight\, functions}\quad 
w(x).
\label{Eq.3.8}
\end{equation}
The method of weighted residuals necessitates that a weighted integral 
of the residual vanishes. In the Galerkin method the choice of weight functions
is $w_p(x)\equiv N_p(x)$, where $N_p(x)$ are shape functions that define 
the FEM ansatz of the solution. As we have shown in
Ref.~\cite{PVR.97}, the choice $w_p(r)=r^2 N_p(r)$ 
produces symmetric stiffness matrices for the radial equations.
Using the standard representation for the Pauli matrices, the radial
relativistic Hartree-Bogoliubov equations (\ref{equ.2.22}) can be written
in matrix form
\begin{eqnarray}
\label{equ.3.3}
\Bigl[(\partial_r+r^{-1})\,\sigma_3\otimes\sigma_3\sigma_1-
\kappa r^{-1}\,\sigma_3\otimes\sigma_1+
m\,\sigma_3\otimes(\sigma_3-{\bf 1}_2)+
S(r)\,\sigma_3\otimes\sigma_3 +\nonumber\\
V_0(r)\,\sigma_3\otimes {\bf 1}_2+
\sigma_1\otimes {\bf 1}_2\hat\Delta (r)-
\lambda\,\sigma_3\otimes{\bf 1}_2\Bigr]\Psi(r) -
E\,{\bf 1}_4\Psi(r) = 0
\end{eqnarray}
Eq. (\ref{equ.3.3}) represents a system of
four coupled integro-differential equations.
For the matrices which define the structure of these equations
we use the notation
\begin{eqnarray}
\label{equ.3.4}
{\bf A}_1:=\sigma_3\otimes\sigma_3\sigma_1, &
{\bf A}_2:=\sigma_3\otimes\sigma_1, &
{\bf A}_3:=\sigma_3\otimes(\sigma_3-{\bf 1}_2),\nonumber \\ 
{\bf A}_4:=\sigma_3\otimes\sigma_3, &
{\bf A}_5:=\sigma_3\otimes{\bf 1}_2,& 
{\bf A}_6:=\sigma_1\otimes{\bf 1}_2 ;
\end{eqnarray}
The method of weighted residuals transforms the Dirac-Hartree-Bogoliubov equations 
into a finite system of algebraic equations.
The system forms a generalized eigenvalue problem 
${\bf A}\, X =\varepsilon\, {\bf N}\, X$ with global stiffness 
matrices
\begin{eqnarray}
\label{equ.3.5.a}
{\bf A}=\bigl< w_{p'}\big\vert
(\partial_r+r^{-1})\, {\bf A}_1-
\kappa r^{-1}\,{\bf A}_2+
m\, {\bf A}_3 -
\lambda\, {\bf A}_5+
S(r)\, {\bf A}_4 +\nonumber\\
V_0(r)\, {\bf A}_5+ 
{\bf A}_6\,\hat\Delta (r)\big\vert N_p\bigr>
\end{eqnarray}
and
\begin{equation}
\label{equ.3.5.b}
{\bf N}=\bigl< w_{p'}\big\vert {\bf 1}_4 \big\vert N_p\bigr>.
\end{equation}
The number of equations in the system 
is thus four times the number of nodes on the finite element mesh.
The eigenvalue matrix equation reads
\begin{eqnarray}
\label{equ.3.6}
\Bigl[
{\bf S}_1\otimes {\bf A}_1+{\bf S}_2\otimes {\bf A}_1 -
\kappa {\bf S}_2\otimes {\bf A}_2 + m {\bf S}_3\otimes {\bf A}_3
+ {\bf S}_4\otimes {\bf A}_4 + {\bf S}_5\otimes {\bf A}_5 + {\bf S}_6\otimes {\bf A}_6
-\lambda {\bf S}_3\otimes {\bf A}_5\Bigr] x \nonumber \\
= 
\varepsilon \Bigl[ {\bf S}_3\otimes{\bf 1}_4\Bigr]x
\end{eqnarray}
where $x$ is a vector with components $(x^{(g_U)}_1,x^{(f_U)}_1,x^{(g_V)}_1,
x^{(f_V)}_1,...,x^{(g_U)}_n,x^{(f_U)}_n,x^{(g_V)}_n,x^{(f_V)}_n)^T$.
The global stiffness matrices ${\bf S}_1$ to ${\bf S}_6$ correspond to 
the various operator terms in Eq.
(\ref{equ.3.3}) 
\begin{eqnarray}
\label{Eq.3.9}
{\bf S}_1 &=& 
\Bigl[ \bigl< w_{p'}(r)\big\vert \partial_r \big\vert N_p(r)\bigr>\Bigr], 
\nonumber\\
{\bf S}_2 &=& 
\Bigl[\bigl< w_{p'}(r)\big\vert r^{-1} \big\vert N_p(r) \bigr>\Bigr],
\nonumber\\
{\bf S}_3 &=& \Bigl[\bigl< w_{p'}(r) \big\vert N_p(r)\bigr>\Bigr], 
\nonumber\\
{\bf S}_4 &=& 
\Bigl[\bigl< w_{p'}(r) \big\vert S(r) \big\vert N_p(r)\bigr>\Bigr], 
\nonumber\\
{\bf S}_5 &=& 
\Bigl[\bigl< w_{p'}(r) \big\vert V(r)\big\vert N_p(r)\bigr>\Bigr],
\nonumber\\
{\bf S}_6 &=& 
\Bigl[\bigl< w_{p'}(r') \big\vert\Delta(r',r)r^2\big\vert N_p(r)\bigr>\Bigr]. 
\end{eqnarray}
Details of the calculation and construction of the matrices ${\bf S}_1$ to
${\bf S}_5$ have been described in Appendix A of Ref.~\cite{PVR.97}.
The matrix elements of ${\bf S}_6$ have
to be calculated for the nonlocal pairing interaction $\Delta(r',r)$.
Using the same notation as in Ref.~\cite{PVR.97}, 
the matrix elements of the local stiffness matrix
in the reference element representation are
\begin{eqnarray}
\label{Eq.3.10}
{\bf S}_6^{\rm loc}=
\Bigl[\bigl<N_{q'}\big\vert r'^2\Delta(r',r)r^2\big\vert N_q\bigr>\Bigr]
= \qquad\qquad\qquad\qquad\qquad     \nonumber\\
\Bigl[ m^2\Delta\zeta^2 a^6\int\limits_0^1d\rho'\int\limits_0^1d\rho
N_{q'}(\rho')(\zeta_{p'}+\Delta\zeta\rho')^{3m-1}                 \nonumber \\
\Delta(a(\zeta_{p'}+\Delta\zeta\rho')^m,a(\zeta_p+\Delta\zeta\rho)^m)
(\zeta_p+\Delta\zeta\rho)^{3m-1}N_q(\rho) \Bigr]
\end{eqnarray}
\noindent
For the FEM discretization of the Klein-Gordon equations  
we use the ansatz
\begin{eqnarray}
\label{Eq.3.9.a}
\sigma (r)=\sum\limits_p \sigma_p  N_p(r) \, \\
\label{Eq.3.9.b}
\omega^0 (r)=\sum\limits_p \omega^0_p  N_p(r) \, \\
\label{Eq.3.9.c}
\rho^0 (r)=\sum\limits_p \rho^0_p  N_p(r) \, \\
\label{Eq.3.9.d}
A^0 (r)=\sum\limits_p A^0_p  N_p(r).
\end{eqnarray}
where the node variables $\sigma_p,$ $\omega^0_p,$ $\rho^0_p$ and
$A^0_p$ correspond to field amplitudes at the mesh point p.
For the Klein-Gordon equations we use the same type of
shape functions $N_p(r)$, and the same mesh as in the 
FEM discretization of the RHB equation (\ref{equ.3.3}). 
In this way we obtain the following algebraic equations for the
mean field amplitudes of the meson fields
\begin{eqnarray}
\label{Eq.3.10.a}
\Bigl[S^{\sigma}_1+l(l+1)\, S^{\sigma}_2+
m_{\sigma}^2\,  S^{\sigma}_3+S^{\sigma}_4\Bigr]\, {\vec\sigma}\,
&=& {\vec r^{\rm (s)}} \, \\
\label{Eq.3.10.b}
\Bigl[S^{\omega}_1+l(l+1)\, S^{\omega}_2+
m_{\omega}^2\, S^{\omega}_3\Bigr]\,{\vec\omega^0}  & = &
{\vec r^{\rm (v)}} \, \\
\label{Eq.3.10.c}
\Bigl[S^{\rho}_1+l(l+1)\, S^{\rho}_2+
m_{\rho}^2\, S^{\rho}_3\Bigr]\,{\vec\rho^0} & = &
{\vec r^{\rm (3)}}  \, \\
\label{Eq.3.10.d}
\Bigl[S^{A^0}_1+l(l+1)\, S^{A^0}_2\Bigr]\,{\vec A^0} & = &
{\vec r^{\rm (em)}} .
\end{eqnarray}
The node variables $\sigma_p,$ $\omega^0_p,$ $\rho^0_p$ and
$A^0_p$ are grouped 
into the vectors ${\vec\sigma}=(\sigma_1,...,\sigma_n)^T$,
${\vec\omega^{\,0}}=(\omega_1,...,\omega_n)^T$, 
${\vec\rho^{\,0}}=(\rho^0_1,...,\rho^0_n)^T$, ${\vec A^0}=(A^0_1,...,A^0_n)^T$,
and \begin{eqnarray}
\label{Eq.4.10.a}
S_1^{\sigma} = S_1^{\omega} = S_1^{\rho} = S_1^{A} &=&
\bigl< w_{p'}(r)\big\vert \partial_r^2 + 2\,r^{-1}\,\partial_r 
\big\vert N_p(r) \bigr>,                                       \, \\
\label{Eq.4.10.b}
S_2^{\sigma} = S_2^{\omega} = S_2^{\rho} = S_2^{A} &=&
\bigl< w_{p'}(r)\big\vert r^{-2}\big\vert N_p(r)\bigr>,      \, \\
\label{Eq.4.10.c}
S_3^{\sigma} = S_3^{\omega} = S_3^{\rho} = S_3^{A} &=&
\bigl< w_{p'}(r)\big\vert N_p(r)\bigr>,           \,  \\
\label{Eq.4.10.d}
S_4^{\sigma} &=& 
\bigl< w_{p'}(r)\big\vert g_2\sigma(r)+g_3\sigma(r)^2\Big\vert N_p(r)\bigr>.
\end{eqnarray}
The components of the right hand side vectors are defined as
\begin{eqnarray}
r^{\rm (s)}_{p'}&=&-g_{\sigma}\bigl<w_{p'}(r)\big\vert\rho_{\rm s}(r)\bigr>, \nonumber\\
r^{\rm (v)}_{p'}&=& g_{\omega}\bigl<w_{p'}(r)\big\vert\rho_{\rm v}(r)\bigr>, \nonumber\\
r^{(3)}_{p'}&=& g_{\rho}\bigl<w_{p'}(r)\big\vert\rho_3(r)\bigr>, \nonumber\\
r^{\rm (em)}_{p'}&=& e \bigl<w_{p'}(r)\big\vert\rho_{\rm em}(r)\bigr>.
\label{Eq.4.11}
\end{eqnarray}
\noindent
Details on the calculation of the stiffness matrices 
${\bf S}^{\sigma}_1$,${\bf S}^{\sigma}_2$,${\bf S}^{\sigma}_3$,
${\bf S}^{\sigma}_4$,${\bf S}^{\omega}_1$,
${\bf S}^{\omega}_2$,${\bf S}^{\omega}_3$,${\bf S}^{\rho}_1$,${\bf S}^{\rho}_2$,
${\bf S}^{\rho}_3$,
${\bf S}^{A^0}_1$,${\bf S}^{A^0}_2$, and the right-hand side vectors 
${\vec r^{\rm (s)}},{\vec r^{\rm (v)}},{\vec r^{\rm (3)}},
{\vec r^{\rm (em)}}$, are given in Appendix B of Ref.~\cite{PVR.97}.
%
%
%
%
%
Next we discuss the structure and occupation pattern
of the global stiffness matrices of the RHB equation and describe
the inclusion of boundary conditions.
The boundary conditions for 
the spherical symmetric case are
\begin{eqnarray}
\label{equ.4.11}
\matrix{
f^{(U)}(r=0)=0 & \mbox{and} & f^{(V)}(r=0)=0 & for & \kappa=-1 \cr
g^{(U)}(r=0)=0 & \mbox{and} & g^{(V)}(r=0)=0 & for & \kappa=+1 \cr
g^{(U)}(r=0)=0 & \mbox{and} & f^{(U)}(r=0)=0 & for & \vert\kappa\vert>1 \cr
g^{(V)}(r=0)=0 & \mbox{and} & f^{(V)}(r=0)=0 & for & \vert\kappa\vert>1 }
\end{eqnarray}
%
%
%
%
%
%
%
%
\noindent
and
\begin{eqnarray}
\label{equ.4.12}
\matrix{ g^{(U)}(r_{\rm max})=0 & \mbox{and} & g^{(V)}(r_{\rm max})=0 }
\quad\mbox{\rm for all}\quad \kappa.
\end{eqnarray}
In Figs. 3 (a) and (b) we display the occupation pattern of the global 
stiffness matrics $A$ and $N$, respectively. The occupation pattern 
corresponds to first order finite
elements (linear shape functions). Shaded squares indicate the 
positions of non-zero matrix elements.
The occupation pattern displays a dominant block diagonal band
structure. This structure results from the contributions of the local
terms in (\ref{equ.3.3}). Each diagonal block consists of several
$4\times 4$-blocks. The number of $4\times 4$-blocks 
is determined by the order of finite elements:
$(n_{\rm ord}+1)^2$. From Eq.(\ref{equ.3.6}) we notice that 
the occupation pattern in a
$4\times 4$-block results from a superposition of occupation
patterns of the matrices $A_k$ (\ref{equ.3.4}). Non-zero matrix elements
outside the  block diagonal band structure result from nonlocal terms
that correspond to the finite range pairing interaction
(dark shaded squares in Fig. 3 (a)). 
In the Figs. 3 (c) and (d), and Figs. 3 (e) and (f) we display the 
occupation patterns for the choice of 
second and third order finite elements, respectively. 
To each pair of nodes $(p,p')$ on the finite element mesh, there 
%
%
%
%
%
\noindent
corresponds
a $4\times 4$-block $\bigl<w_p\vert\hat H\vert N_{p'}\bigr>$, which has
to be multiplied by a vector $X_{p'}$. Depending on the
boundary conditions (\ref{equ.4.11}), one or more componets of
$X_{p'}$ are set to zero. Boundary conditions are taken into account
by simply eliminating columns and rows with the corresponding index from
both stiffness matrices ${\bf A}$ and ${\bf N}$. We illustrate this procedure in Figs.
3 (a) and (b) for the case $\kappa = -1$
(boundaries $r=0$ and $r=r_{\rm max}$). In the Figs. 3 (c) and (d) 
the boundary conditions correspond to
$\kappa = +1$, while in the Figs. 3 (e) and (f) boundary conditions
for the case $\vert\kappa\vert > 1$ are displayed. 
%
%
%
%
%
%
%
%
%
%
%
%
%
%

%
\section {An illustrative calculation}
%
%
In this section we present an illustrative calculation for 
the ground state of the spherical nucleus $^{124}$Sn.
The self-consistent calculation is performed for the 
mean-field parameter
set NL1\cite{RRM.86}, and the D1S~\cite{BGG.84} parameters for the
finite range pairing interaction (\ref{equ.2.18}).
The $Z=50$ protons in Sn form a closed shell. Single-particle 
wave functions of proton states are calculated as solutions of the
radial Dirac equation, as described in Ref.~\cite{PVR.97}. $N=74$
neutrons in the isotope $^{124}$Sn partially occupy the major 
shell $N=50 - 82$. For neutrons 
we solve the radial RHB-equations (\ref{equ.2.22}), and Klein-Gordon
equations (\ref{equ.2.23.a}) to (\ref{equ.2.23.d}) for the meson and 
photon fields. 
In the initial step of the self-consistent iteration,
Woods-Saxon shapes are used
\begin{eqnarray}
\label{Eq.6.1.a}
S(r)=S(0)\Bigl(1+\exp{({{r-r_s}\over a})}\Bigr)^{-1},  \, \\
\label{Eq.6.1.b}
V(r)=V(0)\Bigl(1+\exp{({{r-r_s}\over a})}\Bigr)^{-1}.
\end{eqnarray}
for the scalar $\sigma$ and
vector $\omega$ potentials.
The contribution of the $\rho$-meson to the effective
potential is set to zero in the first iteration step.
The initial Coulomb potential
corresponds to a homogenous spherical charge 
%
%
%
%
%
%
%
%
\noindent
distribution of
radius $r_s$. 
For $^{124}$Sn 
the initial values of the scalar and
vector potentials at $r=0$ are  chosen
$S(0)=-395\,\mbox{\rm MeV}$ and $V(0)=320\,\mbox{\rm MeV}$, respectively.
$a=0.5\, {\rm fm}$ and
$r_{\rm s}=6.0\, {\rm fm}$.
The RHB equations are solved for $\kappa = \pm1$, ...$\pm 7$. 
The inclusion of additional $\kappa$-blocks 
in the self-consistent calculation did not change the results for 
ground state properties. 
As is illustrated in the energy diagram in Fig. 2, 
the diagonalization of the resulting eigenvalue problem is performed
in a window of positive quasi-particle energies
$[0, 100]\,{\rm MeV}$. 

\noindent In each step of the self-consistent iteration, the value of the 
chemical potential $\lambda$ has to be adjusted in such a way 
that the expectation value of the particle number operator 
equals the number of neutrons, i.e. 74 in this case.  The correct 
value of $\lambda$ is found as the root of the function
%
%
%
%
%
%
%
%
%
\begin{equation}
\label{equ.4.1}
dn(\lambda):=N-\sum\limits_{\kappa}\sum\limits_{n}
\int\limits_0^{\infty } dr r^2(g^{(V)}_{n,\kappa}(\lambda,r)^2+
                             f^{(V)}_{n,\kappa}(\lambda,r)^2)
\end{equation}
where N is the number of neutrons, and the second term is 
the trace of the density matrix. In the initial steps of the 
self-consistent iteration, $\lambda$
is not calculated with a very high precision. The precision 
increases with the number of 
iteration steps, i.e. with the accuracy with which the 
mean-field is calculated.
If $\lambda^{(i)}_0$ is the calculated value of the chemical
potential in the i-th iteration step, 
we define an intervall
$I^{(i+1)}:=[\lambda^{(i)}_0-\Delta\lambda^{(i)},
             \lambda^{(i)}_0+\Delta\lambda^{(i)}]$ in which
the new value $\lambda_0^{(i+1)}$ is to be found.
The width of the interval is defined
%
%
%
%
%
%
%
%
%
%
\begin{equation}
\label{equ.4.2}
\Delta\lambda^{(i)}={{\Delta\lambda^{(0)}}\over{(i+1)^n}}
\end{equation}
where $\Delta\lambda^{(0)}$ is an input parameter, and we take $n=2$.
This procedure leads to good convergence, and accurate values of 
$\lambda$ are obtained when the iterations approach the self-consistent 
solution. After the first few RHB iterations, the number of nested 
 $\lambda$-iteration steps varies
between 1 and 3. 

In what follows we present results for the self-consistent solution that
correspond to the ground state of $^{124}$Sn.
The meson fields have been calculated with precision
$10^{-3}\,{\rm MeV}$, and the accuracy for single quasi-particle 
energies is $10^{-4}\,{\rm MeV}$. In Figs. 4a, 4b,
4c, and 4d we display the normalized amplitudes of the quasi-particle
spinors. 
%
%
%
%
%
%
%
%
%
\noindent
The normalization for quasi-particle states is defined: 
\begin{equation}
\label{equ.4.3}
1 = \int\limits d^3r( \Phi_U^{\dagger}({\bf r})\Phi_U({\bf r}) +
                      \Phi_V^{\dagger}({\bf r})\Phi_V({\bf r}) )
\end{equation}
\noindent
In Fig. 5 we display the baryon densities (\ref{equ.2.3.f})
calculated with 150 linear shape functions in a radial box of 30 fm.
The dashed curve corresponds to the density calculated in the first iteration step
with the initial Woods-Saxon potentials, and the solid curve is the
the self-consistent result. 
%
%
%
%
%
\noindent
Since the densities are very sensitive to the details 
of the numerical approximations, we use them to
choose the most effective cut-off in the stiffness matrices of the
Hartree-Bogoliubov equations. Namely, as explained in 
Section 5, we only construct stiffness matrices with 
band diagonal structure. The width of the bands is an input parameter, 
i.e. the cut-off parameter of the matrices.
A too small value of the cut-off parameter means that many off-diagonal
matrix elements of the pairing interaction are neglected. 
A large cut-off makes the diagonalization
of the algebraic eigenvalue equations slow. In the calculation of the
baryon density in Fig. 5 we have used a cut-off value of 50. 
The resulting global stiffness matrices have a band diagonal structure of width
101. This value is too small, and unphysical oscillations
of the density are observed. 
The amplitude of these
oscillations corresponds to the values of the largest pairing matrix
elements that have been neglected. Most of our calculations, and in
particular the results that follow, have been performed with 
a cut-off parameter 60.

In Fig. 6 we show, for each value of the $\kappa$-quantum number, 
the contributions of individual quasi-particle states 
to the pairing field. We plot the quantities
\begin{eqnarray}
\label{equ.4.4}
\pmatrix{ \Delta^{(g)}_V(r) \cr \Delta^{(f)}_V(r) } =
\int\limits_0^{r_{max}}dr'r'^2
\pmatrix{ \Delta^{(gg)}(r,r') & 0 \cr
           0 & \Delta^{(ff)}(r,r') }
\pmatrix{ g^{(V)}_{n\kappa}(r') \cr
          f^{(V)}_{n\kappa}(r') }.
\end{eqnarray}
%
%
%
%
%
%
\noindent
The contributions to the pairing field are mainly concentrated
on the the surface of the nucleus. 
In Fig. 7 we also display the integrated kernel of the integral operator
of the pairing interaction
\begin{eqnarray}
\label{equ.4.5}
\pmatrix{ \Delta^{(g)}(r) \cr \Delta^{(f)}(r) } =
\int\limits_0^{r_{max}}dr'r'^2
\pmatrix{ \Delta^{(gg)}(r,r') & 0 \cr
           0 & \Delta^{(ff)}(r,r') } \pmatrix{ 1 \cr 1 }
\end{eqnarray}
%
%
%
%
%
%
\noindent
In order to illustrate how the single-particle density and pair matrices depend
on the size $r_{\rm max}$ of the spherical box, in Fig. 8 we 
display the quantities $N_{n\kappa}$ and 
$P_{n\kappa}$~\cite{DFT.84}
\begin{eqnarray}
\label{equ.4.6.a}
N_{n\kappa}=
\int\limits_0^{\rm max}dr r^2( g^{(V)}_{n\kappa}(r)g^{(V)}_{n\kappa}(r)+
f^{(V)}_{n\kappa}(r)f^{(V)}_{n\kappa}(r)), \\
\label{equ.4.6.b}
P_{n\kappa}=
-\int\limits_0^{\rm max}dr r^2( g^{(U)}_{n\kappa}(r)g^{(V)}_{n\kappa}(r)+
f^{(U)}_{n\kappa}(r)f^{(V)}_{n\kappa}(r)), 
\end{eqnarray}
which constitute a measure of the contribution of the $n$-th quasi-particle
state to the density matrix $\rho$ and to the pairing tensor
$\kappa$, respectively. 
$(2j+1)N_n$ is the contribution to the particle number.
In Figs. 8a and 8b we plot 
$N_n$ and $P_n$ as functions of quasi-particle energy for the
$s_{1/2}$-states ($\kappa=-1$). Self-consistent results are shown 
for three sizes of the radial box: $r_{\rm max}=15,\,30$ and $40\,{\rm fm}$.
The precision of the calculations is:
$10^{-3}\,{\rm MeV}$ for the meson fields, and $10^{-4}\,{\rm MeV}$ for
the quasi-particle energies. The largest contributions to the density $\rho$ 
come from the three quasi-particle states at $1.52\,{\rm MeV}$, 
$26.18\,{\rm MeV}$ and
$51.82\,{\rm MeV}$). The state at $1.52\,{\rm MeV}$, which is closest 
to the Fermi level $\lambda$, gives the largest contribution to the 
pairing tensor. For $\kappa=-1$ the contribution
to the pairing comes from these three peaks that represent bound states.
Both quantities, $N_n$ and $P_n$, asymptotically decay with increasing
quasi-particle energy.
Therefore, an extension of the energy window above 
$100\,{\rm MeV}$ would 
not change the calculated quantities.
Similar results are obtained for other values of $\kappa$.
%
%
%
%
%
%
\noindent
In Tables 1 and 2 we illustrate the numerical accuracy with which 
calculations can be performed. From 80 to 200 linear finite elements have
been used in the calculations. 14 values of $\kappa$ 
are included: $\kappa = -7$ to $\kappa = +7$. 
In Table 1 quantities that 
characterize the bulk properties of 
$^{124}$Sn are displayed. 
For all quantities we observe convergence with the increase 
of the size of the radial box, and 
of the quasi-particle energy window. In addition, in the right column,
we list the
corresponding values calculated with a computer program that uses an
expansion in oscillator basis functions~\cite{LEL.96}. For a nucleus 
that is still far away from the drip line, the finite element
discretization and the oscillator basis expansion should produce
very similar results. For drip line nuclei we expect the finite
element method to provide more accurate solutions.  
$r_{\rm max}$ and $E_{\rm max}$ denote the size of the radial box
and of the energy window, respectively.
The energies ${\bf E}_p$ and ${\bf E}_n$
are defined as
\begin{equation}
\label{equ.4.7}
{\bf E}_{n(p)} = \sum\limits_{i} 2\vert\kappa_i\vert v_{i,n(p)}^2 E_{i,n(p)},
\end{equation}
where the $E_{i,n(p)}$ denotes the canonical energies, and $v_{i,n(p)}$ the
occupation probabilities. The pairing field $\Delta_{aa}(r,r')$ and 
the pairing tensor $\kappa_{aa}(r,r')$ are used to calculate 
the pairing energy ${\bf E}_{\rm pair}$ 
\begin{equation}
\label{equ.4.8}
E_{pair}=\int\limits_0^{\infty}dr r^2\int\limits_0^{\infty}dr' r'^2
\sum_{a}\kappa_{aa}(r,r')\Delta_{aa}(r,r').
\end{equation}
The sum in (\ref{equ.4.8}) runs over all values of $\kappa$.
In addition we display in Table 1 the total energy of the $\sigma$-field
${\bf E}_{\rm sig}$, the contribution of nonlinear terms to the 
$\sigma$-energy ${\bf E}_{\rm nl}$, the energy of the $\omega$-field
${\bf E}_{\rm ome}$, the energy of the $\rho$-field 
${\bf E}_{\rm rho}$, the Coulomb energy 
${\bf E}_{\rm pho}$, the total binding energy ${\bf E}_{\rm b}$
and the binding energy per nucleon ${{\bf E}_{\rm b}}\over {\bf A}$.
These quantities are defined in Refs. \cite{PVR.97} and \cite{GRT.89}.
The mean square radii ${\bf rms}$ are defined
\begin{equation}
\label{equ.4.9}
{\bf rms}\,=\, 
\sqrt{ { { \int\limits_0^{\infty}dr r^4
\sum\limits_{i} 2\vert\kappa_i\vert
\Phi_{V,i}^{\dagger}(r)\Phi_{V,i}(r) } 
\over
{ \int\limits_0^{\infty}dr r^2
\sum\limits_{i} 2\vert\kappa_i\vert
\Phi_{V,i}^{\dagger}(r)\Phi_{V,i}(r) } } }
\end{equation}
In Table 2 we compare results for canonical energies and occupation
probabilities of single-neutron states in $^{124}$Sn.
%
%
\section{Program structure}
%
The program is coded in C++. The implementation of the relativistic
mean field model in the Hartree approximation for spherical doubly-closed
shell nuclei has been described in Ref.~
\cite{PVR.97}. In this section we only describe the changes that have 
been made in order to extend the program to open-shell nuclei, and
include pairing correlations in the framework of relativistic 
Hartree-Bogoliubov theory. 

The main part of the program consists of seven classes:
{\bf MathPar}: numerical parameters used in the code,
{\bf PhysPar}: physical parameters (masses,
coupling constants, etc.),
{\bf FinEl}: finite elements, {\bf Mesh}:
mesh in coordinate space,
{\bf Nucleon}: neutrons and protons in
the nuclear system, {\bf Meson}: mesons and photon with correponding
mean fields and the Coulomb field, and the class {\bf LinBCGOp}.
A detailed description of these classes can be found in Ref.~\cite{PVR.97}.
In the following we describe additional methods and parameters that are 
used in the present version of the program. The program allows
one type of nucleons (protons or neutrons) to partially occupy a
major shell of the nuclear shell model. A partially occupied major 
shell is referred to as an open shell. Nuclei with both proton and
neutron shells open are generally deformed. The program is restricted 
to nuclear systems with spherical symmetry. For the type of 
nucleons that occupy open shells the relativistic Hartree Bogoliubov 
(RHB) equations have to be solved for a set of $\kappa$-quantum numbers:
$\pm 1$, $\pm 2$, $\pm 3$, ... The class
{\bf MathPar} contains the new parameter {\it Kapa\_max} which equals 
the maximal absolute value of $\kappa$.
Other new parameters are: {\it Mesh\_rmax\_mes} for
the size of the mesh on which the meson field equations are solved,
{\it Nucleon\_maxrhbst[PhysPar::NumNucTypes]} for the maximal number of
quasi-particle states, {\it Lambda0} for the initial value of the
chemical potential, {\it D\_Lamb0 } for the initial step in
the $\lambda$-iteration procedure, and {\it Numb\_rhb\_tol} for the initial
tolerance in the calculation of the number of particles for a particular 
value of $\lambda$. Due to nonconservation of the number of particles,
solutions of RHB equations in general do not correspond to the
correct number of nucleons. In the $\lambda$-iteration the value of the
chemical potential is adjusted in such a way that the expectation value 
of the particle number operator (trace of the density matrix), equals
the number of nucleons for the specific nucleus.
 
The set of physical parameters of the nucleus, contained
in the class {\bf PhysPar}, is extended by the parameter
{\it RHB\_nuc\_number[PhysPar::NumNucTypes]}, which is the number of 
nucleons for which the RHB equations are solved. 
The prototype method {\it set\_nucleus\_XX()} can be used to define 
the numbers of protons and neutrons.
In addition to the mean field parameter sets NL1, NL2, NLSH and NL3, 
the class {\bf PhysPar}contains two parameter sets for 
the Gogny interactio: D1~\cite{DG.80} and D1S~\cite{BGG.84}.

Essential changes have been made in the class {\bf nucleon}. The method
{\it solve( Meson const{\rm \&} \_sigma, Meson const{\rm \&} \_omega,
                     Meson const{\rm \&} \_rho, Meson const{\rm \&} \_photon )}
distinguishes between nucleons that occupy closed shells, for
which the Dirac equation has to be solved, and nucleons in open shells,
for which the wave functions are solutions of Hartree-Bogoliubov 
equations. A new constructor
{\it make\_gog\_stiff(int nvec,int ikap,int *iab,int *iz,double* A,double *N)}
for stiffness matrices of the RHB equations
has been added to the class. {\it make\_gog\_stiff} is used only in the first
step of the $\lambda$-iteration, for each $\kappa$.
The method 
{\it store\_gog\_stiff(int nvec,int ndx,int ikap,int* iab,int* iz,
double* A, double *N)}.
stores the resulting
stiffness matrices 
in the private fields $int** siz$, $int** siab$, $double** SA$,
and $double** SB$. 
If the diagonalization has to be repeated just for a different value 
of the chemical potential,
the reconstruction of the stiffness matrices is performed by 
{\it read\_gog\_stiff(nvec,ndx,ikap,iab,iz,A,N)} and
{\it make\_new\_lamb\_gog\_stiff(A,lambda,lamb\_old)}.
Matrix elements that depend on $\lambda$ are replaced with new values.
\hfill\break
{\it boucond(kapa,nvec,n\_eig,iab,iz,A,N,{\rm \&}nvec2,{\rm \&}n2)} includes 
boundary conditions in the stiffness matrices. 
Rows and columns are eliminated, and the resulting matrices are written 
in condensed form.

The largest matrix elements of the finite range pairing interaction are
concentrated near the diagonal of the stiffness matrix. The absolute values
of the pairing matrix elements monotonically decrease as one goes 
away from the diagonal,
and the outermost are many orders of magnitude smaller
than those at the diagonal. Therefore, elements of the stiffness matrices
are only calculated in blocks within a band diagonal structure. The 
{\it mat\_cutoff(int nvec2,int n2,int* iz,int* iab,double *A,double* N,
int* nvec3,int* ndf3)} removes from the global stiffness matrix
all elements which lie outside a band of chosen width, 
and performes a condensation of the matix.
The eigenvalue problem with band matrix structure is solved by the
bisection method {\it bisec(n2,A,N,iab,iz,en\_low\_rhb,en\_upp\_rhb,tol,
lam,XX,ndx,mdx,ndf3)}. 
Eigenvalues and eigenvectors are calculated in a window \hfill\break 
$\left[ low\_rhb; upp\_rhb \right]$ of the energy parameter.

The function {\it elim\_spur\_rhb(mdx,XX,lam,ist,{\rm \&}numb\_lev,kapa)}
eliminates spurious  solutions
from the spectrum produced by {\it bisec}.
{\it add\_dens\_rhb} adds contributions to the source terms of
the boson fields. For a calculation performed with some 
value of the chemical potential,
the method {\it trace\_dens(int ist,int numb\_lev,int kapa)}
calculates the trace of the density matrix for each $\kappa$.
A combination of the secant method and the bisection method
is used by 
{\it calc\_new\_lambda\_secant(lamb\_iter,dn,{\rm \&}la\_l,{\rm \&}la\_r,{\rm \&}dn\_l,{\rm \&}dn\_r,
lam\_old)} to calculate new $\lambda$ values. In the following table we 
list all the new member functions which have been included in the 
class {\bf nucleon}.
\vspace{1 cm}
\hfill\break
{\bf List of new member functions in class nucleon:} \hfill\break 
\hfill\break 
    void   write\_canon\_energies( double** vv, double** Ec );   \hfill\break
    void   write\_solutions(int kapa,double *lam,double **XX,int mdx );   
\hfill\break
    void   write\_rhb\_states(int k);   \hfill\break
    void   write\_canonical\_basis(int kapa,int n,double **CG,double **CF);   
\hfill\break
    void   write\_dens\_kap(int k);   \hfill\break
    void   write\_stiff\_matels(int k,int nvec,int n,int *iz,int *iab,
                              double *A,double *B);   \hfill\break
    void   calc\_canonical\_basis();   \hfill\break
    void   calc\_NnPn();   \hfill\break
    void   scan\_delta\_wave(int k);   \hfill\break
    void   scan\_delta();   \hfill\break
    double energy\_pair();   \hfill\break
    double energy\_pair\_delta();   \hfill\break
    double radius\_ms();   \hfill\break
    double sum\_r2();   \hfill\break
    double norm();   \hfill\break
    double fac(double n);   \hfill\break
    void       make\_kapa\_list();   \hfill\break
    double     guvwave( double const* g, int kapa, int ife, int iga ) const;
\hfill\break
    double     fuvwave( double const* f, int kapa, int ife, int iga ) const;   
\hfill\break
    void       make\_stiff\_delta\_force(int lamb\_iter,int nvec,int ikap,
                                      int *iab,int *iz,
                                      double *ast,double *bst);   \hfill\break
    void       make\_gog\_stiff(int nvec,int k,int *iab,int *iz,
                              double *ast,double *bst);   \hfill\break
    void       make\_new\_lamb\_gog\_stiff(double *ast,double lamb\_new,
                                       double lamb\_old);   \hfill\break
    void       store\_gog\_stiff(int nvec,int n,int ikap,int *iab,int *iz,
                               double *A,double *B);   \hfill\break
    void       read\_gog\_stiff(int nvec,int n,int ikap,int *iab,int *iz,
                              double *A,double *B);   \hfill\break
    void       boucond(int kapa,int nvec,int n,int *iab,int *iz,
                       double *ast,double *bst,int *nvec2,int *n2);   \hfill\break
    void       mat\_cutoff(int nvec,int n,int *iz,int *iab,
                          double *ast,double *bst,int *nvec3,int *ndf3);  
\hfill\break
    void       calc\_gofac();   \hfill\break
    void       make\_s3\_locst( int ife, double** s3 );   \hfill\break    
    void       make\_loc\_gog\_st(int ikap,int ife1,int ife2,double** s6,
                                                          double** s7);   
\hfill\break
    void       make\_loc\_stiff\_delta(int ikap,int ife1,double **s6,
                                                      double** s7);   \hfill\break
    double     delta(int swgf,int k1,int ife1,int iga1,int ife2,int iga2);   
\hfill\break
    double     delta\_delt(int swgf,int k1,int ife1,int iga1);   \hfill\break
    void       elim\_spur\_rhb(int mdx,double** XX,double const* lam,
                             int ist,int *numb\_lev,int kapa );   \hfill\break
    void       pickup\_rhb\_state(int iw,int kapa,double const* X,double lam);
\hfill\break
    void       setup\_stiffmats();   \hfill\break
    void       copy\_new\_old();   \hfill\break
    void       calc\_dens0\_rhb();   \hfill\break 
    double     density\_rhb(int ife,int iga);   \hfill\break
    void       make\_dens\_matr(double *A,double *B,int *iab,int *iz,
                              int cutoff,int *nvec,int *nn);   \hfill\break
    void       make\_kk\_dens\_matr(int k,double *A,double *B,int *iab,int *iz,
                                 int *nvec,int *n);   \hfill\break    
    void       make\_dens\_matr\_sdiag(int k,double **A,int *n);   \hfill\break
    void       add\_dens\_rhb(int ist,int numb\_lev,int kapa);   \hfill\break
    void       large\_r\_dens(double r\_cutoff);   \hfill\break
    void       bisec\_lambda(int lamb\_iter,double dn,
                            double *la\_l,double *la\_r,
                            double *dn\_l,double *dn\_r);   \hfill\break
    void       bisec\_lambda(int lamb\_iter,double dn,double *dn1,double *dn2);
\hfill\break
    void       calc\_new\_lambda\_BCS();   \hfill\break
    void       calc\_new\_lambda\_secant(int lamb\_iter,double dn,
                                      double *la\_l,double *la\_r,
                                      double *dn\_l,double *dn\_r,
                                      double *lam\_old);   \hfill\break    
    double     calc\_nfe\_mes();   \hfill\break
    double     pair\_tens\_kap(int swgf,int k,int ife1,int iga1,
                                            int ife2,int iga2);   \hfill\break
    double     trace\_dens(int ist,int numb\_lev,int kapa);   \hfill\break
    double     delta\_lambda(double dn,double sum);   \hfill\break
    double     add\_sum(int ist,int numb\_lev,int kapa);   \hfill\break
    double     gofac0(double l1,double l2,double j1,double j2,double lam);   
\hfill\break
    double     gofac1(double l1,double l2,double j1,double j2,double lam);   
\hfill\break
    double     pl(int l, double x);   \hfill\break   
    double     vl(int l,int ife1,int iga1,int ife2,int iga2,double mu\_gog);
\hfill\break
    double     clebsh\_gordon\_coeff(double j1,double m1,
                                   double j2,double m2,
                                   double j3,double m3);   \hfill\break
    double     wig\_3j(double j1, double j2, double j3,
                      double m1, double m2, double m3);   \hfill\break
    double     wig\_6j(double j1, double j2, double j3,
                      double l1, double l2, double l3);   \hfill\break
    double     del(double j1, double j2, double j3);      
\vspace{0.4cm}
\hfill\break 
\noindent
For the meson fields there is a possibility to choose the size 
of the finite element mesh different from that used in the 
solution of the Hartree-Bogoliubov equations. In some cases 
we have found that, by taking a smaller radial box for the 
meson fields, better numerical accuracy is obtained for
quasi-particle spinors that correspond to nucleon states in the 
continuum.
The number of finite elements in the meson mesh is determined
by the member function {\it double Meson::calc\_nfe\_mes() }
in the constructor of the class {\bf meson}.
The size parameter of the radial box {\it r\_max\_mes} has been 
included in the class {\bf mesh}.
In {\it numutils.cc} we have included a new
solver for matrix diagonalization 
{\it void sdiag(int n, double **a, double *d, double **x, double *e, int is) }.
It is based
on the Householder algorithm, and is used for matrices of
dimension smaller than $1000\times1000$. {\it sdiag}
diagonalizes the extremely ill conditioned density matrices 
$\rho_{kk'}$ in the member function
{\it calc\_canonical\_basis} of class {\bf nucleon}.
The diagonalization is performed in the final transformation
of single quasi-particle solutions to the canonical basis of single
particle states.
\newpage
\appendix
%
%

\section{RHB equations for nuclear systems with spherical symmetry}
\vskip0.5cm
%
The coordinate transformation 
\begin{eqnarray}
\label{equ.A.1}
{\cal T}:\, \left[\left. 0,\infty \right)\right.\otimes 
            [0,\pi)\otimes [0,2\pi)& \longrightarrow & {\bf R}^3
\nonumber \\
     (r,\theta,\phi) & \longmapsto & (r\,\sin{\theta}\,\cos{\phi},
                                      r\,\sin{\theta}\,\sin{\phi},
                                      r\,\cos{\theta} )
\end{eqnarray}
defines the RHB equation (\ref{equ.2.2}) in 
spherical coordiantes $(r,\theta,\phi)$. 
\begin{eqnarray}
\label{equ.A.2}
\hat H_D = -i\,\bigl( \alpha_3\,\partial_r +
\alpha_1\,{1\over r}\,\partial_{\theta} + 
\alpha_2\, {1\over{ r\sin{\theta} }}\,\partial_{\phi} \bigr)+
V(r) + m\,\beta + S(r)\,\beta 
\end{eqnarray}
is the Dirac hamiltonian in spherical coordinates.
The standard representation is used for the matrices $\alpha_i\, 
(i=1,2,3)$: $\alpha_i = \sigma_1\otimes\sigma_i$, where 
$\sigma_i$ are the Pauli matrices.
For a system with spherical symmetry, the scalar and the
vector potential read 
\begin{eqnarray}
\label{equ.A.3}
\hat S(r)&=& g_{\sigma}\,\sigma(r) \nonumber\\
\hat V(r)&=& g_{\omega}\,\omega^0(r)+g_{\rho}\,\vec\tau\,\vec\rho(\vec r)
+e\,{{(1-\tau_3)}\over 2}\,A^0(r)
\end{eqnarray}
Single-particle eigenfunctions of the Dirac hamiltonian $\hat H_D$ 
are at the same time eigenvectors of the total angular momentum 
$\hat J$, its z-component $\hat J_z$, and of the operator 
\begin{equation}
\label{equ.A.4}
\hat K := -i\beta\bigl( \Sigma_2\partial_{\theta} - {1\over{\sin{\theta}}}
\Sigma_1\partial_{\phi} \bigr),
\end{equation}
where $\Sigma_i:={\bf 1}_2\otimes\sigma_i\,(i=1,2,3)$.
For the eigenfunctions of $\hat H_D$
\begin{equation}
\label{equ.A.5}
\hat K\, \Psi_{\kappa} = -\kappa\, \Psi_{\kappa} \qquad 
(\kappa =\pm 1,\pm 2,\pm 3,... ),
\end{equation}
and this relation motivates the ansatz 
\begin{equation}
\label{equ.A.6}
\Psi^T_{n,\kappa,m}(r,\theta,\phi):=(g_{n\kappa}(r),i\,f_{n,-\kappa}(r))^T
\otimes (\Omega_{\kappa,m}(\theta,\phi),\Omega_{-\kappa,m}(\theta,\phi))^T.
\end{equation}
The Dirac hamiltonian takes the form
\begin{equation}
\label{equ.A.7}
\hat H_D = 
-i\,\bigl( \alpha_3\,\partial_r + \gamma_3\, r^{-1}\, \hat K\bigr)
+\hat V(r) + m\,(\beta -{\bf 1}_4) + \hat S(r)\,\beta, 
\end{equation}
and the left hand side operator of the RHB equation (\ref{equ.2.2})
for a spherical symmetric system reads
\begin{equation}
\label{equ.A.8}
\hat H_{RHB}=\sigma_3\otimes
\bigl[ -i\,\bigl( \alpha_3\partial_r + \gamma_3{1\over r}\hat K\bigr)
+\hat V(r) + m\beta + \hat S(r)\beta \bigr] 
-\lambda \sigma_3\otimes{\bf 1}_4 
+ \sigma_1\otimes \hat\Delta (r).
\end{equation}
The ansatz for the nucleon spinor in (\ref{equ.2.2}) is
\begin{eqnarray}
\label{equ.A.9}
\pmatrix{ U \cr V} = 
\pmatrix{ g^{(U)}(r)\,\Omega_{\kappa,m}(\theta,\phi) \cr
       i\,f^{(U)}(r)\,\Omega_{-\kappa,m}(\theta,\phi)\cr
          g^{(V)}(r)\,\Omega_{\kappa,m}(\theta,\phi) \cr
       i\,f^{(V)}(r)\,\Omega_{-\kappa,m}(\theta,\phi) }
\end{eqnarray}
where the radial functions $g(r)$ and $f(r)$ depend on the
principal quantum number $n$, and on $\kappa$. The pairing tensor is
defined $\kappa_{cd}(\vec r,\vec r') := U_c^*(\vec r)V_d^T(\vec r')$,
and we postulate 
\begin{eqnarray}
\label{A.10}
\kappa_{nn'\tilde\kappa\tilde\kappa'mm'}(r,r',\theta,\theta',\phi,\phi')= 
\qquad\qquad\qquad\qquad\qquad\qquad\qquad\qquad
\nonumber \\
\nonumber \\
\pmatrix{ g^{(U)}_{n\tilde\kappa}(r)g^{(V)}_{n'\tilde\kappa'}(r')
\Omega^*_{\tilde\kappa \tilde m}(\theta,\phi)
\Omega^T_{\tilde\kappa' \tilde m'}(\theta',\phi') &0\cr
0 &f^{(U)}_{n\tilde\kappa}(r)f^{(V)}_{n'\tilde\kappa'}(r')
\Omega^*_{-\tilde\kappa \tilde m}(\theta,\phi)
\Omega^T_{-\tilde\kappa' \tilde m'}(\theta',\phi') }.
\end{eqnarray}
For a pairing interaction of finite range, the kernel $\Delta_{ab}({\bf r},{\bf r'})$
(\ref{equ.2.5},)is written in the form   
\begin{eqnarray}
\label{A.11}
\Delta_{\kappa\kappa'}(r,\theta,\phi,r',\theta',\phi') = \nonumber\\
{1\over 2}
\sum\limits_{\tilde n,\tilde n'}\sum\limits_{\tilde\kappa,\tilde\kappa'}
\sum\limits_{\tilde m,\tilde m'}
V_{\kappa\kappa'\tilde\kappa\tilde\kappa'}(r,r')
\delta_{\tilde n,\tilde n'}\delta_{\tilde\kappa\tilde\kappa'}
\delta_{\tilde m\tilde m'}
\kappa_{\tilde n\tilde n'\tilde\kappa\tilde\kappa'\tilde m\tilde m'}
(r,r',\theta,\theta',\phi,\phi'),
\end{eqnarray}
and the integral operator (\ref{equ.2.4})of the pairing field, as a function
of spherical coordinates
\begin{eqnarray}
\label{A.13}
\hat\Delta(r,\theta,\phi)=                        
\int\limits_0^{\infty}\int\limits_0^{\pi}
\int\limits_0^{2\pi} dr' r'^2 d\theta\sin{\theta}d\phi
{1\over 2}\sum_{\tilde\kappa}V_{\kappa\kappa\tilde\kappa\tilde\kappa}
 \kappa_{\tilde\kappa\tilde\kappa}(r,\theta,\phi,r',\theta',\phi').
\end{eqnarray}
%
%
%
%
%
\section{Two-nucleon matrix elements of the Gogny interaction}
\vskip0.5cm
%
%
For isospin T=1 pairing, i.e. pairing interaction between identical
nucleons, the Gogny force (\ref{equ.2.18}) can be written in the form
\begin{equation}
\label{equ.B.1}
V(1,2)=\sum\limits_{i=1,2}V_i(\vert {\bf r}_1-{\bf r}_2\vert )~~
(A_i+B_iP^{\sigma})
\end{equation}
where the spin operator acting on the two-nucleon state is defined
\begin{eqnarray}
\label{equ.B.2}
P^{\sigma} \bigl.\vert ({1\over 2}{1\over 2})S\bigr> = 
(2S-1)\bigl.\vert({1\over 2}{1\over 2})S\bigr> =
\left\{ \matrix{ +1 & \mbox{\rm for} & S=1; \cr
                -1 & \mbox{\rm for} & S=0; } \right.
\end{eqnarray}
The radial interaction
\begin{equation}
\label{equ.B.3}
V_i(\vert {\bf r}_1-{\bf r}_2\vert ) = 
e^{-{ { ( {\bf r}_1-{\bf r}_2)^2 } \over {\mu_i^2}} }
\end{equation}
can be written as an infinite sum of terms, each separable in the
angular coordinates of the two nucleons
\begin{equation}
\label{B.4}
V(\vert{\bf r}_1 -{\bf r}_2\vert)=\sum\limits_{\lambda=0}^{\infty}
V_{\lambda}(r_1,r_2)P_{\lambda}(\cos{\theta_{12}}),
\end{equation}
where
\begin{equation}
\label{B.5}
P_{\lambda}(\cos{\theta_{12}})={{4\pi}\over{2\lambda +1}}
\sum_{\mu}Y_{\lambda\mu}^*(\hat {\bf r}_1)Y_{\lambda\mu}(\hat {\bf r}_2),
\end{equation}
and
\begin{equation}
\label{B.15}
V_{\lambda }(r_1,r_2)={{2\lambda +1}\over 2}\int\limits_{-1}^1
d\cos{\theta_{12}}V(\vert{\bf r}_1 -{\bf r}_2\vert)
P_{\lambda}(\cos{ \theta_{12}} )
\end{equation}
The non-antisymmetrized, two-nucleon matrix element between 
non-normalized states reads
\begin{equation}
V_{JM} = \bigl<rlj, r'l'j'\big\vert V(\vert{\bf r} -{\bf r'}\vert)
\bigl.\big\vert r\tilde l \tilde j, r' \tilde l' \tilde j'\bigr>_{JM},
\end{equation}
where $J$ and $M$ are the total angular momentum of the two-nucleon state,
and its $z$-projection, respectively. Since the interaction contains 
the spin operator, it is convenient to calculate the matrix element in 
the LS-coupling scheme. The transformation between the $jj$ and 
LS-coupling schemes for the two nucleon state
\begin{eqnarray}
\label{B.100}
\bigl.\big\vert (l,l'),jj; JM\bigr>=
\sum\limits_{LS}\hat j\hat j'\hat L\hat S
\left\{ \matrix{ {1\over 2} & {1\over 2}
& S \cr l & l' & L \cr j & j' & J }\right\}
\bigl.\big\vert (ll')SL;JM\bigr>
\end{eqnarray}
where L is the total orbital angular momentum, and
S the total spin of the two-nucleon state. We use the 
short-hand notation $\hat j:=\sqrt{ 2j+1}$. The matrix 
element can then be written 
\begin{eqnarray}
\label{B.200}
V_{JM}=\sum\limits_{LS}\sum\limits_{\tilde L,\tilde S}
\hat j\hat j'\hat L \hat S\hat{\tilde j}\hat{\tilde j'}\hat{\tilde L}
\hat {\tilde S}\bigl< rr',(ll')SLJM\vert V(\vert{\bf r} -{\bf r'}\vert)\vert
\tilde r\tilde r',(\tilde l\tilde l')\tilde S\tilde L JM\bigr>
\left\{ \matrix{ {1\over 2} & {1\over 2}
& S \cr l & l' & L \cr j & j' & J }\right\}
\left\{ \matrix{ {1\over 2} & {1\over 2} & \tilde S \cr 
                 \tilde l & \tilde l' & \tilde L \cr 
                 \tilde j & \tilde j' & J } \right\}.
\end{eqnarray}
Using the Slater expansion (\ref{B.4}), the LS-coupling matrix element 
takes the form
\begin{equation}
\label{B.6}
V^{JM}_{S\tilde S L\tilde L} =
\sum\limits_{\lambda}{ {4\pi}\over {2\lambda +1} }
V_{\lambda}(r,r')
\sum\limits_{\mu}
\bigl<(ll')SLJM\vert Y^*_{\lambda\mu}(\hat {\bf r})
Y_{\lambda\mu}(\hat{\bf r'})(A+B\hat P^{\sigma})\vert
(\tilde l\tilde l')\tilde S\tilde L JM\bigr>,
\end{equation}
where the radial factors $V_{\lambda}(r,r')$ are defined in 
(\ref{B.15}), and the angular part is calculated
\begin{eqnarray}
\label{B.11}
&\sum\limits_{\mu}
\bigl<(ll')SLJM\vert Y^*_{\lambda\mu}(\hat {\bf r})
Y_{\lambda\mu}(\hat{\bf r'})(A+B\hat P^{\sigma})\vert
(\tilde l\tilde l')\tilde S\tilde L JM\bigr>=\nonumber \\
&\delta_{S\tilde S}\delta_{L\tilde L}(A+B(2S-1))
 (-1)^{L+l+\tilde l}~~
{{\hat l\hat l'\hat {\tilde l}\hat {\tilde l'}\hat \lambda^2}\over {4\pi}} 
\pmatrix{ l & \lambda & \tilde l \cr 0 & 0 & 0}
\pmatrix{ l' & \lambda & \tilde l' \cr 0 & 0 & 0}
\left\{ \matrix{ \tilde l & \tilde l' & L \cr l' & l & \lambda} \right\}
\end{eqnarray}
For the pairing interaction we only consider matrix elements between 
two-nucleon states with total angular momentum $J=0$. In this case 
the $9j$-coefficients in (\ref{B.200}) reduce to  $6j$-coefficients, 
and the two-nucleon matrix element reads
\begin{eqnarray}
\label{B.14}
&V_{J=0}=\delta_{jj'}\delta_{\tilde j\tilde j'}
\delta_{ll'}\delta_{\tilde l\tilde l'}
(2l+1)(2\tilde l +1)
\hat {j}\hat {\tilde j'} \nonumber \\
&\sum\limits_{S=0,1}(-1)^{j+\tilde j +1 +S} (2S+1)
\left\{ \matrix{  l &  l & S \cr 
                 {1\over 2} & {1\over 2} & j} \right\}
\left\{ \matrix{ \tilde l & \tilde l & S \cr 
                 {1\over 2} & {1\over 2} & \tilde j} \right\}\nonumber \\
&\sum\limits_{\lambda}
V_{\lambda}(r,r')( A+B(2S-1) )
\pmatrix{ l & \lambda & \tilde l \cr 0 & 0 & 0}^2 
\left\{ \matrix{ l &  \tilde l & \lambda \cr 
                 \tilde l & l & S} \right\}.
\end{eqnarray}
The allowed integer values for $\lambda$ in the sum are 
\begin{equation}
\vert l-\tilde l\vert \leq \lambda \leq l+\tilde l
\end{equation}
The matrix elements essentially consist of two terms:
$S=0$ and $S=1$. Introducing the explicit expressions for the 
$3j$ and $6j$-coefficients, the two terms read
\begin{eqnarray}
&V_{J=0~S=0}= (A-B) \delta_{jj'}\delta_{\tilde j\tilde j'}
\delta_{ll'}\delta_{\tilde l\tilde l'} \hat {j}\hat {\tilde j} 
\sum\limits_{\lambda {\rm even}} (-1)^{\lambda} V_{\lambda}(r,r')\nonumber \\
&{{ (2g-2l)! (2g-2\lambda)! (2g-2 \tilde l)!}\over {2(2g+1)!}}
\left [ {{g!}\over {(g-l)! (g-\lambda)! (g-\tilde l)!}}\right ]^2
\end{eqnarray}
and
\begin{eqnarray}
&V_{J=0~S=1} = \frac{1}{2} (A+B) \delta_{jj'}\delta_{\tilde j\tilde j'}
\delta_{ll'}\delta_{\tilde l\tilde l'} \hat {j}\hat {\tilde j}
{{(\frac{3}{4} + l(l+1) - j(j+1))(\frac{3}{4} + \tilde l(\tilde l +1) 
- \tilde j(\tilde j+1))}\over {l(l+1) \tilde l(\tilde l +1)}}
\sum\limits_{\lambda {\rm even}} (-1)^{\lambda} V_{\lambda}(r,r') \nonumber \\
&\left [ l(l+1) + \tilde l(\tilde l +1) + \lambda (\lambda +1) \right ]
{{ (2g-2l)! (2g-2\lambda)! (2g-2 \tilde l)!}\over {2(2g+1)!}}
\left [ {{g!}\over{(g-l)! (g-\lambda)! (g-\tilde l)!}}\right ]^2
\end{eqnarray}
where $2g = l+ \lambda + \tilde l$.



\newpage
\leftline{\Large {\bf Figure Captions}}
\parindent = 2 true cm
\begin{description}
\item[Fig. 1] 
Vector space homomorphisms between the general single-particle basis, 
the basis of quasi-particle states, and the canonical basis of single-particle
states. The diagrams illustrate the transformations 
(\ref{equ.2.5.a}) to (\ref{equ.2.5.c}). 

\item[Fig. 2]
Relativistic Hartree-Bogoliubov model for a finite nucleus. Single-particle
eigenspectrum of a Dirac hamiltonian (left and center), and single 
quasi-particle eigenspectrum 
of the Hartree-Bogoliubov equations (right).

\item[Fig. 3a]
Occupation pattern of the global stiffness matrix ${\bf A}$
(\ref{equ.3.5.a}) for $\kappa = -1$ and linear
shape functions. The dark grey squares indicate matrix elements
of the finite range Gogny interaction. Due to the nonlocal character
of the interaction, the matrix elements
are distributed 
over the whole matrix. The light grey squares correspond to matrix elements
which result from the local
operators in (\ref{equ.2.22}). They form a block diagonal band
structure. Boundary conditions are taken into account by eliminating  
the corresponding rows and columns. The matrix is
symmetric and we display the global index of nonzero matrix elements, as used 
by the bisection method in the solution of the eigenvalue problem. 

\item[Fig. 3b]
Occupation pattern of 
the overlap matrix ${\bf N}$ (\ref{equ.3.5.b}), for $\kappa = -1$ and linear
shape functions.

\item[Fig. 3c]
Same as in Fig. 3a, but for $\kappa = +1$ and quadratic shape functions.

\item[Fig. 3d]
Same as in Fig. 3b, but for $\kappa = +1$ and quadratic shape functions.

\item[Fig. 3e]
Same as in Fig. 3a, but for $\vert\kappa\vert > 1$ and cubic shape functions.

\item[Fig. 3f]
Same as in Fig. 3b, but for $\vert\kappa\vert > 1$ and cubic shape functions.

\item[Fig. 4a] 
Self-consistent normalized quasi-particle wave functions.
Radial amplitudes of U-components (upper left), and V-components (upper right),
for the $1s_{1/2}$
(solid), $2s_{1/2}$ (dashed), and $3s_{1/2}$ (doted) bound states in $^{124}$Sn. 
The radial amplitudes of continuum $s_{1/2}$-states are shown in the lower
part of the figure: U-components (left) and V-components (right).
The self-consistent solution is 
calculated with 100 linear shape functions on a uniform radial mesh 
[$0,30\,{\rm fm}$].

\item[Fig. 4b] 
Same as in figure 4a, but for $p_{1/2}$-waves.

\item[Fig. 4c] 
Same as in figure 4a, but for $p_{3/2}$-waves.

\item[Fig. 4d] 
Same as in figure 4a, but for $h_{11/2}$-waves.

\item[Fig. 5]
Baryon density of $^{124}$Sn after the first iteration step (dashed),
and for the self-consistent solution (solid). The results correspond to
a calculation with 100 linear shape functions on a uniform radial mesh 
[$0,30\,{\rm fm}$].

\item[Fig. 6]
$\kappa = \pm 1,\pm 2, \pm 3$ contributions to the pairing field.
For each value of $\kappa$ we display the largest contributions from 
individual states. The quantities that we plot are defined in 
(\ref {equ.4.4}).

\item[Fig. 7]
The components $\Delta^{(g)}(r)$ and $\Delta^{(f)}(r)$ of
the self-consistent 
pairing field (\ref {equ.4.4}), calculated on a radial mesh 
of $15\,{\rm fm}$ (solid) and $30\,{\rm fm}$ (dot dashed).
The pairing field is concentrated on the 
surface of the nucleus. 

\item[Fig. 8]
Logarithmic plots of
$N_n$ (\ref{equ.4.6.a}) and $P_n$  
(\ref{equ.4.6.b})), as functions of quasi-particle energy for the
$s_{1/2}$-states. Results are displayed for three sizes of the radial box:
$r_{\rm max}=15\, {\rm fm}$ (dashed), $r_{\rm max}=30\,{\rm fm}$
(dot dashed) and $r_{\rm max}=40\,{\rm fm}$
(solid).

\end{description}
\newpage        
%
%
%
%
\begin{table}[H]
\begin{center}
\begin{tabular}{|cc|c|c|c|c|c|c|c|c|c|c|}
\hline
    &   &    &       &       &      &      &
    &                             \\
{\bf test run}& & 1&  2    &   3   &  4   &  5   &  6  
    &  7                \\
    &   &FEM &FEM    &FEM    &FEM   & FEM  &FEM
       &  Osc. basis   \\ \hline
    &   &    &       &       &      &     &
           &                     \\

${\bf r_{\rm max}}$ &[fm]& 15.00  & 20.00 & 30.00 & 40.00  &     
  20.00 & 20.00 &             \\  
${\bf E_{\rm max}}$ &[MeV]& 100.00 & 100.00 & 100.00 & 100.00 &   
  70.00  & 110.00  &             \\  
${\bf E_{\rm p}}$   &[MeV]& -1320.74  & -1320.68 & -1320.79&-1320.79&
     -1320.51 & -1320.80 &             \\  
${\bf E_{\rm n}}$   &[MeV]& -1719.42 & -1646.68 &-1719.56&-1719.56&
     -1719.17  & -1719.54 &             \\  
${\bf E_{\rm pair}}$&[MeV]& -19.948 & -19.9494 &-19.9510&-19.9342&
     -19.5666 & -19.9901 & -19.68  \\  
${\bf E_{\rm sig}}$ &[MeV]& 17126.70 & 17126.20 &17126.20&17126.2&
     17125.30 & 17126.40 & 17432.33   \\   
${\bf E_{\rm nl}}$  &[MeV]& -324.004 & -323.988 &-323.983&-323.983&
     -324.019  &-323.977 &  -327.910     \\   
${\bf E_{\rm ome}}$ &[MeV]& -14699.40 &  -14699.27 & -14699.10 &-14699.00&
 -14698.10  & -14699.20 &  -14674.06   \\   
${\bf E_{\rm rho}}$ &[MeV]& -55.460 & -55.455 &-55.448 &-55.448&
     -55.449 & -55.449  & -55.529     \\   
${\bf E_{\rm pho}}$ &[MeV]& -366.072 & -366.072 &-366.073 &-366.073&
     -366.066 &-366.075 & -365.71     \\   
${\bf E_{\rm tot}}$ &[MeV]& -1060.58 & -1060.69  &-1060.83&-1060.82&
    -1055.37 &-1060.86 & 1051.96     \\   
${\bf E}_b/A$ & [MeV]& -8.553 &  -8.554 & -8.555 & -8.555&
     -8.550  &-8.555 & -8.48       \\  
${\bf\lambda }$ & [MeV]& -6.65727 & -6.65718  & -6.65706 & -6.65668 & 
    -6.65719 &-6.65729& -6.65786          \\  
${\bf rms_{\rm neutron}}$ & [fm] & 4.90853& 4.90869& 4.90874 &4.90874 &
     4.90825 & 4.90871 &  4.91073    \\   
${\bf rms_{\rm proton}}$ & [fm] & 4.59741&4.59745& 4.59751 &4.59741 &
     4.59738  &4.59740 &  4.60187    \\   
${\bf rms_{\rm nucleus}}$ &[fm]&4.78555&4.78561& 4.78564 &4.78564 &
     4.78515 &4.78562 & 4.78859     \\   \hline
\end{tabular}
\end{center}
\vskip 0.4cm
{\small {\bf Table 1} \quad
Bulk properties of $^{124}$Sn. Calculated quantities are compared 
for various sizes of the radial mesh, and with results of an 
expansion in the oscillator basis~\cite{LEL.96}.
}
\end{table}

\vspace{0.4cm}
%
%
%
%
%
\begin{table}[H]
\begin{center}
\begin{tabular}{|cc|c|c|c|c|c|c|c|c|c|}
\hline
    &       &       &       &      &      &
           &                      \\
{\bf test run} & & 1&  2    &   3   &  4   &  5   &  6  
                   \\
    &       &       &       &      &      &
    &           \\ \hline
    &       &       &       &      &     &
           &             \\
${\bf r_{\rm max}}$ & [fm] & 15.00  &  20.00 & 30.00 & 40.00  &
  20.00 & 20.00              \\  

${\bf E_{\rm max}}$ &[MeV]& 100.00 & 100.00 & 100.00 & 100.00 &   
  70.00& 110.00             \\  

${\bf E_{\rm 1s_{1/2}}}$ &[MeV]&-51.834& -51.834&-51.834&-51.836&
     -51.828 & -51.832     \\  

${\bf E_{\rm 2s_{1/2}}}$ &[MeV]&-39.294&-39.298&-39.303&-39.300&
     -39.290  & -39.305     \\  

${\bf E_{\rm 3s_{1/2}}}$ &[MeV]&-8.227&-8.231 &-8.235&-8.235&
     -8.221& -8.235     \\  

${\bf E_{\rm 4s_{1/2}}}$ &[MeV]&12.988&12.997 &13.050&12.902&
     12.981& 13.085   \\  

${\bf E_{\rm 1p_{3/2}}}$ &[MeV]&-43.078&-43.076&-43.075&-43.077&
    -43.069& -43.067  \\  

${\bf E_{\rm 2p_{3/2}}}$ &[MeV]&-27.023&-27.029&-27.035&-27.032&
    -27.018& -27.043 \\  

${\bf E_{\rm 3p_{3/2}}}$ &[MeV]&2.977&2.969  &2.961&2.961&
     2.983 & 2.965   \\  
${\bf E_{\rm 1p_{1/2}}}$ &[MeV]&-42.976&-42.998&-43.006&-43.008&
     -42.923 & -43.001    \\  

${\bf E_{\rm 2p_{1/2}}}$ &[MeV]&-24.499&-24.489&-24.481&-24.479 &
     -24.498 & -24.486    \\

${\bf E_{\rm 3p_{1/2}}}$ &[MeV]&3.863&3.850 &3.843&3.842 &
     3.894  & 3.849     \\

${\bf E_{\rm 1d_{5/2}}}$ &[MeV]&-38.359&-38.358 &-38.357&-38.357 &
     -38.358  &  -38.359     \\  

${\bf E_{\rm 2d_{5/2}}}$ &[MeV]&-10.678&-10.683&-10.686&-10.686&
     -10.671 & -10.685      \\  

${\bf E_{\rm 3d_{5/2}}}$ &[MeV]&13.245&13.265 &13.271&13.218&
     13.203  & 13.283       \\  

${\bf E_{\rm 1d_{3/2}}}$ &[MeV]&-35.690&-35.691 &-35.692&-35.692&
 -35.681   & -35.693  \\  

${\bf E_{\rm 2d_{3/2}}}$ &[MeV]&-8.489&-8.494 &-8.496&-8.497&
     -8.471 & -8.496 \\  

${\bf E_{\rm 3d_{3/2}}}$ &[MeV]& 13.851 &13.887 & 14.051 &13.801&
     13.766  & 14.096  \\  

${\bf E_{\rm 1g_{7/2}}}$ &[MeV]&-16.971&-16.970&-16.969&-16.969&
     -16.998  & -16.969        \\  

${\bf E_{\rm 2g_{7/2}}}$ &[MeV]&11.400&11.512 &11.557&11.551&
     11.289   &  11.466         \\  

${\bf E_{\rm 1h_{11/2}}}$ &[MeV]&-6.349&-6.348 &-6.347&-6.347 &
     -6.355 & -6.347     \\  

${\bf E_{\rm 2h_{11/2}}}$ &[MeV]&22.127&22.116 &22.107&22.110&
     22.138  & 22.212     \\  

    &       &       &       &      &     
           &       &             \\   

${\bf v}^2_{\rm 1s_{1/2}}$ &&0.999995&0.999995&0.999995&0.999995&
     0.999995 & 0.999995  \\  

${\bf v}^2_{\rm 2s_{1/2}}$ &&0.999598&0.999596&0.999594&0.999594&
     0.999611  & 0.999593  \\  

${\bf v}^2_{\rm 3s_{1/2}}$ &&0.984564&0.984573&0.984585&0.984603&
     0.984593  & 0.984558  \\  

${\bf v}^2_{\rm 4s_{1/2}}$ &&0.000048&0.000048&0.000049&0.000049&
     0.000046 & 0.000049  \\  

${\bf v}^2_{\rm 1p_{3/2}}$ &&0.999900&0.999900&0.999899&0.999899&
     0.999835  & 0.999899  \\  

${\bf v}^2_{\rm 2p_{3/2}}$ &&0.998427&0.998419&0.998413&0.998414&
     0.998483  & 0.998410   \\  

${\bf v}^2_{\rm 3p_{3/2}}$ &&0.000499&0.000503&0.000505&0.000504&
     0.000413 & 0.000506    \\  

${\bf v}^2_{\rm 1p_{1/2}}$ &&0.999941&0.999940&0.999939&0.999939&
    0.999964   & 0.999939      \\  

${\bf v}^2_{\rm 2p_{1/2}}$ &&0.999025&0.999019 &0.999015&0.999016&
    0.999082  & 0.999013      \\  

${\bf v}^2_{\rm 3p_{1/2}}$ &&0.000218&0.000219&0.000220&0.000220&
    0.000203 & 0.000221   \\  

${\bf v}^2_{\rm 1d_{5/2}}$ &&0.999205&0.999200&0.999198&0.999198&
    0.999256  & 0.999195  \\  

${\bf v}^2_{\rm 2d_{5/2}}$ &&0.987226&0.987186 &0.987160&0.987166&
     0.987347 & 0.987134  \\  

${\bf v}^2_{\rm 3d_{5/2}}$ &&0.000135&0.000135&0.000136&0.000135&
     0.000128 & 0.000137  \\  

${\bf v}^2_{\rm 1d_{3/2}}$ &&0.999384&0.999381&0.999379&0.999378&
    0.999421 & 0.999376  \\  

${\bf v}^2_{\rm 2d_{3/2}}$ &&0.966344&0.966348&0.966337&0.966350&
     0.966473 & 0.966276  \\  

${\bf v}^2_{\rm 3d_{3/2}}$ &&0.000095&0.000095&0.000096&0.000096&
    0.000077 & 0.000097   \\  

${\bf v}^2_{\rm 1g_{9/2}}$ &&0.992754&0.992711 &0.992680&0.992687&
    0.992934 & 0.992660      \\  

${\bf v}^2_{\rm 2g_{9/2}}$ &&0.000410&0.000415&0.000417&0.000417&
     0.000352 & 0.000419     \\  

${\bf v}^2_{\rm 1h_{11/2}}$ &&0.379037&0.379329 &0.379505&0.379647&
    0.377539   & 0.379604      \\  

${\bf v}^2_{\rm 2h_{11/2}}$ &&0.000127&0.000127&0.000128&0.000128&
    0.000121 & 0.000129  \\  \hline
\end{tabular}
\end{center}
\vskip 0.4cm
{\small {\bf Table 2} \quad
Canonical energies and occupation
probabilities of $\kappa = \pm 1,\pm 2,-3,+5,-6$ 
single-neutron states in $^{124}$Sn.
Self-consistent quantities are compared 
for various sizes of the radial mesh.
}
\end{table}

%

\end{document}